\documentclass[english,prb,floatfix,showpacs,preprint]{revtex4}
\usepackage[T1]{fontenc}
\usepackage[latin1]{inputenc}
\usepackage{array}
\usepackage{float}
\usepackage{amsmath}
\usepackage{graphicx}
\usepackage{amssymb}

\makeatletter

\providecommand{\tabularnewline}{\\}

\usepackage{array}

\usepackage{float}

\makeatletter

\usepackage{array}

\usepackage{float}



\makeatother

\usepackage{babel}
\makeatother
\begin{document}

\title{A large-scale correlated study of linear optical absorption and low-lying
excited states of polyacenes: Pariser-Parr-Pople Hamiltonian}

\author{Priya Sony and Alok Shukla }

\address{Physics Department, Indian Institute of Technology, Powai, Mumbai
400076, INDIA}

\email{psony@phy.iitb.ac.in, shukla@phy.iitb.ac.in}

\begin{abstract}
In this paper we present large-scale correlated calculations of linear
optical absorption spectrum of oligo-acenes containing up to seven
benzene rings. For the calculations we used the Pariser-Parr-Pople
(P-P-P) Hamiltonian, along with the configuration interaction (CI)
technique at various levels such as the full CI (FCI), the quadruple
CI (QCI) and multi-reference singles-doubles CI (MRSDCI). The role
of Coulomb parameters used in the P-P-P Hamiltonian was examined by
considering standard Ohno parameters, as well as a screened set of
parameters. A detailed analysis of the many-body character of the
important excited states contributing to the linear absorption has
also been performed. The results of our calculations have been compared
extensively with the theoretical work of other authors, as well as
with the experiments. 
\end{abstract}


\maketitle

\section{Introduction}

\label{intro}

In last few decades, there has been intensive research in the field
of optical properties of acenes.\cite{gundlach18,nelson-apl72,butko-apl83,tanabe41,kivelson26}
It is well known that with the increase in the number of benzene rings
in the acenes, the HOMO to LUMO gap decreases, making the material
more conducting. It is in fact believed that an infinite linear acene,
\emph{i.e.}, the polyacene could be a metal,\cite{tanabe41} and at
low temperatures, a superconductor.\cite{kivelson26} Thus, higher
acenes can prove to be potential candidates for preparing both optical
and electronic devices. Polyacenes generally crystallize in well-defined
structures, and their crystalline forms have found applications in
novel opto-electronic devices such as light-emitting field-effect
transistors.\cite{fet} Three-dimensional structures of polyacenes
have also been investigated theoretically for their electronic structure,
transport and optical properties by several authors.\cite{silbey,hummer,hummer-prb71}
Oligomers of polyacene such as naphthalene, anthracene, tetracene,
pentacene, \emph{etc}., are materials which are well-known for their
interesting optical properties.\cite{clar} Because of the high symmetry,
they have separate optical response to radiation polarized along the
conjugation direction, vis-a-vis the radiation polarized perpendicular
to it. Several experimental investigations of linear optical properties
of polyacenes have been performed over the years. These include linear
absorption based studies of naphthalene,\cite{platt-naphtha,birks-naphtha,naphtha-anthra-1,berg-naphtha,aleksandrovsky,naphth-2,huebner-naphtha,dick-holhneicher,biermann}
anthracene,\cite{platt-naphtha,naphtha-anthra-1,bergman15,steiner100,lambert81,dick83,lyons4,man-trajmar25,sebastian75,wolf181,sackmann58,biermann}
tetracene,\cite{platt-naphtha,sebastian61,berlman-book,bree-lyons,birk-book,burrow86,dahlberg71,biermann}
pentacene,\cite{platt-naphtha,dahlberg71,sebastian61,penta-2,penta-3,park-apl80,kim-tsf420,puigdollers-tsf427,he-apl84,lee-apl84,biermann}
and hexacene.\cite{hexa-1,biermann} Several theoretical studies of
the low-lying excited states of these materials have also been performed,
such as the early LCAO based study by Coulson,\cite{coulson} LCAO-MO
and perimeter MO approach by Pariser,\cite{pariser24} a free-electron-gas
approach by Platt and Ham \emph{et al}.,\cite{platt,ham} CNDO/2 CI
approach by Hofer \emph{et al}.,\cite{hofer-naphta} \emph{ab initio}
multi-reference M$\varnothing$ller-Plesset (MRMP) theory,\cite{kawashima102}
density-matrix renormalization-group (DMRG) theory using P-P-P model
Hamiltonian,\cite{Ramasesha-1,Ramasesha-2} \emph{ab initio} DFT based
methodologies,\cite{hummer-prb71,grimme,heinze113-hepta,houk-hexa,rubio,wiberg}
self-consistent field-random phase approximation (SCF-RPA) scheme
using P-P-P model Hamiltonian,\cite{baldo77} and CNDO/S2 model parameterization
technique by Lipari \emph{et al.}\cite{lipari-duke63} Among the more
recent studies using the Pariser-Parr-Pople (P-P-P) model Hamiltonian
and many-body methodologies, of Ramasesha and coworkers\cite{Ramasesha-1,Ramasesha-2}
are the foremost.

Most of the theoretical studies mentioned above concentrate either
on a class of excited states of polyacenes, or restrict themselves
to the study of smaller oligomers. In some earlier studies, chiefly
because of the lack of computer power at the time, the level of treatment
of electron-correlation effects was rather modest by contemporary
standards. Additionally, to the best of our knowledge, none of the
earlier theoretical studies reported calculations of optical absorption
spectrum of these materials. Therefore, we believe that there is a
need to perform systematic large-scale many-body calculations of optical
properties of these materials with the following aims in mind: (i)
to compute the linear optical absorption spectra of a range of oligoacenes,
(ii) to understand their evolution with increasing size of the oligomers,
(iii) to understand the influence of electron correlation effects
on them, and finally (iv) to understand the nature of low-lying excited
states contributing to the linear optics. As it is well-known, in
quasi-one-dimensional materials such as conjugated polymers, electron-correlation
effects have profound influence on their optical properties.\cite{sumit-review}
It is with all these issues in mind, in the present work, in a systematic
manner, we undertake a large-scale correlated study of linear optical
absorption in oligoacenes of increasing sizes, namely from naphthalene
to heptacene. The optical absorption spectra have been computed both
for the radiation polarized along the conjugation direction as well
the one polarized perpendicular to it. We have used the P-P-P Hamiltonian
for the purpose, and utilized various configuration interaction (CI)
techniques such as the full CI (FCI), the quadruple CI (QCI), and
the multi-reference singles-doubles CI (MRSDCI) methods. The CI-based
correlation methodology used in the present work is sound, and has
been used successfully by us in the past to study the linear and nonlinear
optical properties of various other conjugated polymers.\cite{shukla2,shukla-ppv,shukla-ppp,shukla-tpa,shukla-thg,sony-pdpa}
Additionally, we have also examined the issue of the influence of
Coulomb parameters on the results by performing calculations with
two distinct sets of parameters, namely the standard Ohno parameters,\cite{ohno}
and a screened set of parameters meant for phenylene-based conjugated
polymers,\cite{chandross} to describe the P-P-P model Hamiltonian.

The remainder of this paper is organized as follows. In section \ref{theory}
we briefly review the theoretical methodology adopted in this work.
In section \ref{results} we present and discuss our results of the
optical absorption spectra of various oligoacenes. Finally, in section
\ref{sec:Conclusions} we summarize our conclusions, and present possible
directions for the future research work.

\section{Theory}

\label{theory}

\begin{figure}
\includegraphics[scale=0.7]{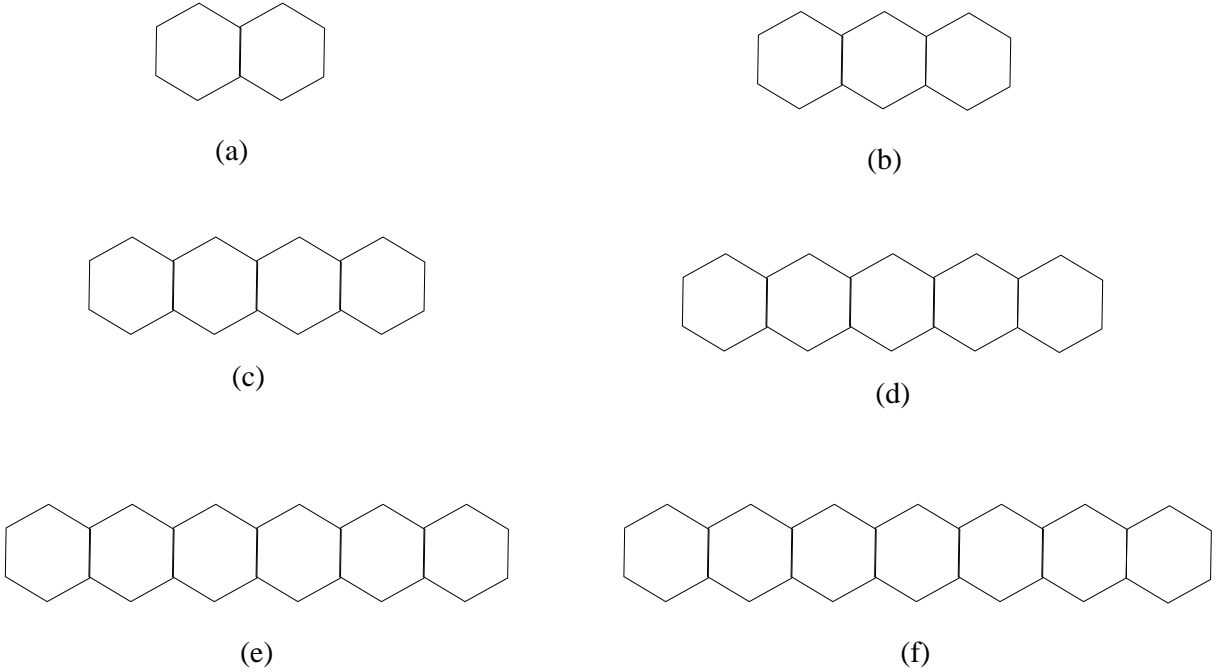}

\caption{Schematic drawings of polyacenes considered in this work, namely,
(a) naphthalene, (b) anthracene, (c) tetracene, (d) pentacene, (e)
hexacene, and (f) heptacene\label{fig-acene}}
\end{figure}

The structures of oligoacenes (C$_{4n+2}$H$_{2n+4}$, $n$=2, 3,
4, 5, 6, and 7) starting from naphthalene up to heptacene are shown
in Fig.\ref{fig-acene}. Oligomers were assumed to lie in the $xy$-plane
with the conjugation direction taken to be along the $x$-axis. They
can be seen as a series of benzene rings fused together, along the
conjugation direction. An alternative way to look at the structure
of oligoacenes is to visualize them as two vertically displaced polyene
chains coupled with each other along the $y$-axis via hopping, and
the Coulomb interactions. From this viewpoint, polyacene is a ladder
like polymer. The point group symmetry of oligoacenes is $D_{2h}$,
so that the one-photon states belong to the irreducible representations
(irreps) $B_{3u}$ or $B_{2u}$, while the ground state belongs to
the irrep $A_{g}$. Also, by convention we assign the ground state
a negative ($-$) particle-hole symmetry. Therefore, by dipole selection
rules, all the optically allowed states must have positive ($+$)
particle-hole symmetry. Thus, the states with negative particle-hole
symmetry will not be visible in the linear optical spectrum. However,
we have calculated the energy of $1B_{3u}^{-}$ state, together with
the optically allowed $1B_{2u}^{+}$ and $1B_{3u}^{+}$ states for
the sake of comparison with the experimental and other theoretical
results. Clar classified the absorption spectra into three bands,
namely, $p$, $\alpha$, and $\beta$.\cite{clar} In our work, transition
to $1B_{2u}^{+}$ state from the $1A_{g}^{-}$ ground state via a
short-axis ($y$-axis) polarized photon corresponds to the $p$ band
($^{1}L_{a}$ band of Platt\cite{platt}). Similarly, transition from
the ground state to $1B_{3u}^{-}$ and $1B_{3u}^{+}$ via a long-axis
($x$-axis) polarized photon corresponds to the $\alpha$ ($^{1}L_{b}$
band of Platt\cite{platt}) and the $\beta$ ($^{1}B_{b}$ band of
Platt\cite{platt}) bands, respectively.

The correlated calculations on the oligoacenes were performed using
the P-P-P model Hamiltonian, which can be written as

\begin{equation}
H=H_{C_{1}}+H_{C_{2}}+H_{C_{1}C_{2}}+H_{ee},\label{eq-ham}\end{equation}
 where $H_{C_{1}}$ and $H_{C_{2}}$ are the one-electron Hamiltonians
for the carbon atoms located on the upper and the lower polyene chains,
respectively. $H_{C_{1}C_{2}}$ is the one-electron hopping between
the two chains, and H$_{ee}$ depicts the electron-electron repulsion.
The individual terms can now be written as,\begin{subequations} \label{allequations}

\begin{equation}
H_{C_{1}}=-t_{0}\sum_{\langle k,k'\rangle}B_{k,k'},\label{eq-h1}\end{equation}
 \begin{equation}
H_{C_{2}}=-t_{0}\sum_{\langle\mu,\nu\rangle}B_{\mu,\nu},\label{eq-h2}\end{equation}
 and

\begin{equation}
H_{C_{1}C_{2}}=-t_{\perp}\sum_{\langle k,\mu\rangle}B_{k,\mu}.\label{eq-h3}\end{equation}
 \end{subequations}

\begin{eqnarray}
H_{ee} & = & U\sum_{i}n_{i\uparrow}n_{i\downarrow}+\frac{1}{2}\sum_{i\neq j}V_{i,j}(n_{i}-1)(n_{j}-1)\textrm{}\label{eq:hee}\end{eqnarray}
 In the equation above, $k$, $k'$ are carbon atoms on the upper
polyene chain, $\mu,\nu$ are carbon atoms located on the lower polyene
chain, while $i$ and $j$ represent all the atoms of the oligomer.
Symbol $\langle...\rangle$ implies nearest neighbors, and $B_{i,j}=\sum_{\sigma}(c_{i,\sigma}^{\dagger}c_{j,\sigma}+h.c.)$.
Matrix elements $t_{0}$, and $t_{\perp}$ depict one-electron hops.
As far as the values of the hopping matrix elements are concerned,
we took $t_{0}=2.4$ eV for both intracell and intercell hopping,
and $t_{\perp}=t_{0}$ consistent with the undimerized ground state
for polyacene argued by Raghu \emph{et al}.\cite{Ramasesha-1} Consequently,
the carbon-carbon bond length has been fixed at $1.4$ \AA , and
all bond angles have been taken to be 120$^{o}$. At this point it
is worthwhile to discuss the issue of the ground state geometry of
oligoacenes. Experimentally speaking, to the best of our knowledge,
the available data on the ground state geometry of various polyacenes
is for the crystalline phase.\cite{campbell-hexa} Therefore, as far
as the ground state geometries of isolated chains are concerned, theoretical
calculations based upon geometry optimization provide very important
input. However, the picture which emerges from such calculations is
far from clear. Raghu \textit{et al}.\cite{Ramasesha-1} studied the
ground-state geometry of polyacenes using a DMRG based approach and
concluded that the Peierls' instability in this polymer is conditional,
with the gain in the electronic energy being proportional to $\delta^{2}$
($\delta$ is the dimerization amplitude), rather than $\delta$ which
is the case for \emph{trans}-polyacetylene. In their next paper, Raghu
\textit{et al}.\cite{Ramasesha-2} concluded that the ground state
geometry of polyacene consists of a weakly distorted structure with
undimerized chains, coupled by slightly longer rungs. Therefore, Raghu
\textit{et al}.\cite{Ramasesha-2} used an undistorted geometry in
their excited state calculations. Houk et al.\cite{houk-hexa}, based
upon their \emph{ab initio} DFT based calculations, had also predicted
a ground state similar to that of Raghu \textit{et al}.\cite{Ramasesha-2}
Several workers have suggested that due to Peierls distortion acenes
possess nonsymmetric geometry,\cite{deleuze,grimme-BL,rubio,wiberg}
but on the other hand Cioslowski\cite{cioslowski} has reported that
at the correlated level, the symmetric geometry is more stable. Also,
Klein and coworkers\cite{klein} studied the distortion of polyacenes
within a many-body valence-bond framework, have suggested that all
the structures (symmetric or nonsymmetric) are close in energy with
symmetric one possessing the lowest energy. Moreover, Niehaus $\textit{et al}.$\cite{niehaus}
have recently studied polyacenes using a tight-binding-based Green's-function
approach, and reported that for $n\leq19$, the symmetric structure
is the most stable. Therefore, keeping these uncertainties in mind,
and the fact that our main aim is to study the influence of electron
correlations on the optical properties of oligoacenes, we have performed
our calculations using the symmetrical structure for oligoacenes.
This choice also allows us to compare our results directly with those
of Ramasesha and coworkers,\cite{ramasesha-soos91,rama-anthra,Ramasesha-2}
who also employed a symmetric geometry in their excited state of oligoacenes
using P-P-P model Hamiltonian.

The Coulomb interactions are parameterized according to the Ohno relationship,\cite{ohno}
\begin{equation}
V_{i,j}=U/\kappa_{i,j}(1+0.6117R_{i,j}^{2})^{1/2}\;\mbox{,}\label{eq-ohno}\end{equation}

where, $\kappa_{i,j}$ depicts the dielectric constant of the system
which can simulate the effects of screening, $U$ is the on-site repulsion
term, and $R_{i,j}$ is the distance in \AA ~ between the $i$-th
carbon and the the $j$-th carbon. In the present work, we have performed
calculations with two parameter sets: (a) {}``standard parameters''
with $U=11.13$ eV and $\kappa_{i,j}=1.0$, and (b) {}``screened
parameters'' with $U=8.0$ eV and $\kappa_{i,j}=2.0$ ($i\neq j)$
and $\kappa_{i,i}=1$, proposed initially by Chandross and Mazumdar
to study phenyl-based conjugated polymers.\cite{chandross} In our
earlier studies of phenyl-based conjugated polymers such as the PDPA,
PPP, and PPV \textit{etc.,} we found that the screened parameters
generally provided much better of description of their optical properties
as compared to the standard ones.\cite{shukla2,shukla-ppv,shukla-ppp,shukla-tpa,shukla-thg,sony-pdpa}

The starting point of the correlated calculations for various oligomers
were the restricted Hartree-Fock (HF) calculations, using the P-P-P
Hamiltonian. The many-body effects beyond HF were computed using different
levels of the configuration interaction (CI) method, namely, full-CI
(FCI), quadruples-CI (QCI), and the multi-reference singles-doubles
CI (MRSDCI). Details of these CI-based many-body procedures have been
presented in our earlier works.\cite{shukla2,shukla-ppv,shukla-tpa,shukla-thg}
From the CI calculations, we obtain the eigenfunctions and eigenvalues
corresponding to the correlated ground and excited states of various
oligomers. The many-body wave functions are used to compute the matrix
elements of the dipole operator connecting the ground state to various
excited states. These dipole matrix elements are in turn used to calculate
the linear optical absorption spectra of various polyacenes.

\section{Calculations and Results}

\label{results}

In this section, first we will briefly discuss the main features of
linear optical spectra of polyacenes computed within the independent-electron
Hückel model. Next we will present and discuss the main results of
this work, namely, the correlated linear absorption spectra of oligoacenes
of increasing sizes, and compare our results with the other available
experimental and theoretical results. A preliminary description of
results for tetracene and pentacene, was presented in an earlier work\cite{priya-acenes-synth}.
However, the results presented here are based upon more extensive
calculations, and, therefore, they supersede our earlier results\cite{priya-acenes-synth}.

\subsection{Hückel Model Results}

\begin{figure}[H]
\begin{centering}
\includegraphics[width=10cm,keepaspectratio]{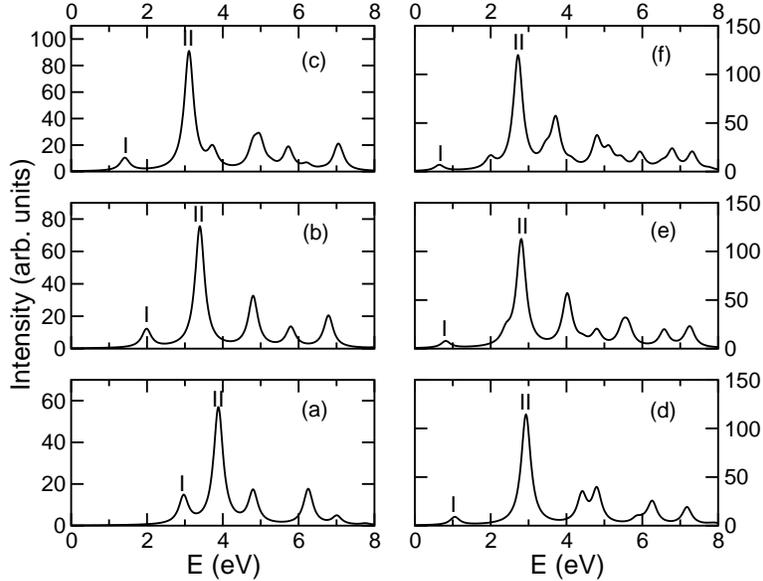}
\par\end{centering}

\caption{Linear optical absorption spectra of (a) naphthalene, (b) anthracene,
(c) tetracene, (d) pentacene, (e) hexacene, and (f) heptacene. In
all the cases a line width of 0.15 eV was assumed.}

\label{fig-huck} 
\end{figure}

Here we briefly discuss the salient features of the linear absorption
spectra of oligoacenes computed using the tight-binding Hückel model,
and presented in Fig.\ref{fig-huck}.

For all the oligomers, the first peak (labeled I) corresponds to $\pi-\pi^{*}$
excitation described by HOMO $(H)\rightarrow$ LUMO $(L)$ transition
through a $y$-polarized photon leading to the $1B_{2u}^{+}$ excited
state of the system. However, with the increasing size of the oligomer,
the intensity of this $H\rightarrow L$ transition decreases, which
is understandable because, in the thermodynamic limit, with the chosen
hopping parameters, the polyacene is metallic. The highest intensity
peak (labeled II) for each of the oligoacene investigated corresponds
to an $x$-polarized photon leading the system to its $1B_{3u}^{+}$
excited state. For oligomers with $n=\mbox{even}$, the highest intensity
peak corresponds to the transitions $H\rightarrow L+n/2$ and $H-n/2\rightarrow L$,
while for those with $n=\mbox{odd}$, the peak II corresponds to transitions
$H\rightarrow L+(n\pm1)/2$ and $H-(n\pm1)/2\rightarrow L$. Moreover,
as the lengths of the oligoacenes increase, as expected, the spectrum
is red shifted due to delocalization of particle-hole pair.

\begin{table}[H]

\caption{The number of reference configurations ($N_{ref}$) and the total
number of configurations ($N_{total}$) involved in the MRSDCI (or
FCI or QCI, where indicated) calculations, for different symmetry
sub-spaces of various oligoacenes.}

\vspace{0.25cm}

\begin{centering}
\begin{tabular}{|c|cc|cc|cc|}
\hline 
Oligomer&
\multicolumn{2}{c|}{$A_{g}$}&
\multicolumn{2}{c|}{$B_{2u}$}&
\multicolumn{2}{c|}{$B_{3u}$}\tabularnewline
&
$N_{ref}$&
$N_{total}$&
$N_{ref}$&
$N_{total}$&
$N_{ref}$&
$N_{total}$\tabularnewline
\hline 
naphthalene&
1$^{a}$&
4936$^{a}$&
1$^{a}$&
4794$^{a}$&
1$^{a}$&
4816$^{a}$\tabularnewline
\hline 
anthracene&
1$^{a}$&
623576$^{a}$&
1$^{a}$&
618478$^{a}$&
1$^{a}$&
620928$^{a}$\tabularnewline
\hline 
tetracene&
1$^{b}$&
193538$^{b}$&
1$^{b}$&
335325$^{b}$&
24$^{c,d}$&
34788$^{c,d}$\tabularnewline
\hline 
pentacene&
1$^{b}$&
1002597$^{b}$&
1$^{b}$&
1707243$^{b}$&
38$^{c}$ &
130196$^{c}$\tabularnewline
&
&
&
&
&
34$^{d}$&
126690$^{d}$\tabularnewline
\hline 
hexacene&
66$^{c}$&
460527$^{c}$&
28$^{c}$&
191944$^{c}$&
54$^{c}$&
393248$^{c}$\tabularnewline
&
40$^{d}$&
299141$^{d}$&
32$^{d}$&
242013$^{d}$&
30$^{d}$&
252420$^{d}$\tabularnewline
\hline 
heptacene&
45$^{c}$&
590599$^{c}$&
30$^{c}$&
415999$^{c}$&
40$^{c,d}$&
653476$^{c,d}$\tabularnewline
&
36$^{d}$&
480032$^{d}$&
22$^{d}$&
270391$^{d}$&
&
\tabularnewline
\hline
\end{tabular}\label{tab:n-ref.}
\par\end{centering}

$^{a}$FCI method with standard as well as screened parameters,

$^{b}$QCI method with standard as well as screened parameters,

$^{c}$using standard parameters,

$^{d}$using screened parameters. 
\end{table}

\subsection{P-P-P Calculations}

Here we present the results of our correlated calculations of linear
absorption on oligoacenes using the P-P-P model Hamiltonian. First
we present and discuss our results for individual acenes, followed
by a unified discussion of their spectra. In Table \ref{tab:n-ref.}
we present the number of reference states ($N_{ref}$) and the dimension
of the Hamiltonian matrix ($N_{total}$) used in our CI calculations
for different symmetry sub-spaces of various oligomers. The fact that
the calculations presented here are quite large-scale is obvious from
$N_{total}$, which, \textit{e.g.}, for pentacene was in excess of
one million for the $A_{g}$ and $B_{2u}$ symmetries. Therefore,
we are confident that our results take into account the influence
of electron-correlation effects quite accurately.

\begin{table}

\caption{Comparison of results of our calculations performed with the standard
(Std.) parameters and the screened (Scd.) parameters with other experimental
and theoretical results for the three most important low-lying states.
For reference 51 (Kadantsev \emph{et al.}),}

results quoted with the asterisk ({*}) correspond to their CISD calculations,
while those without it are their B3LYP results.

\begin{raggedright}
\label{tab:comparion}\begin{tabular}{|c|>{\centering}p{0.8cm}|>{\centering}p{0.8cm}|c|c|c|c|c|c|c|c|c|c|c|c|}
\hline 
\multicolumn{15}{|c|}{~~~~~~Excitation energy (eV)}\tabularnewline
\multicolumn{15}{|l|}{State~Present work~~~~~~~~Experimental~~~~~~~~~~~~~~~~~~~~~~~~~~~~~~~~~~~~~~~~Other
theoretical}\tabularnewline
\multicolumn{15}{|l|}{~~~~~~~~Std. ~Scd.}\tabularnewline
\multicolumn{15}{|l|}{~~~~~~~~para.~para.}\tabularnewline
\hline 
\multicolumn{15}{|c|}{Naphthalene (C$_{\text{10}}$H$_{\text{8}}$)}\tabularnewline
\hline 
$1B_{3u}^{-}$&
3.61&
3.22&
\multicolumn{4}{c|}{3.97\cite{platt-naphtha}, 4.03\cite{biermann}, 4.0\cite{huebner-naphtha}}&
\multicolumn{8}{c|}{4.02\cite{pariser24}, 3.74\cite{ham}, 3.60\cite{ramasesha-soos91},
4.44\cite{rubio}, 4.49\cite{rubio}$^{*}$, 4.21\cite{heinze113-hepta},4.46\cite{grimme},
4.09\cite{hashimoto104}}\tabularnewline
\hline 
$1B_{2u}^{+}$&
4.45&
4.51&
\multicolumn{4}{c|}{4.34\cite{platt-naphtha}, 4.38\cite{biermann}, 4.45\cite{huebner-naphtha},
4.46\cite{aleksandrovsky}}&
\multicolumn{8}{c|}{4.49\cite{pariser24}, 4.54\cite{ham}, 4.46\cite{ramasesha-soos91},
4.35\cite{rubio}, 5.27\cite{rubio}$^{*}$, 4.12\cite{heinze113-hepta},4.88\cite{grimme},
4.62\cite{hashimoto104}}\tabularnewline
\hline 
$1B_{3u}^{+}$&
5.99&
5.30&
\multicolumn{4}{c|}{5.64\cite{platt-naphtha}, 5.62\cite{biermann}, 5.89\cite{huebner-naphtha},
5.95\cite{aleksandrovsky}}&
\multicolumn{8}{c|}{5.94\cite{pariser24}, 5.84\cite{ham}, 5.85\cite{rubio}, 6.24\cite{rubio}$^{*}$,
5.69\cite{heinze113-hepta}, 5.62\cite{hashimoto104}}\tabularnewline
\hline 
\multicolumn{15}{|c|}{Anthracene (C$_{\text{14}}$H$_{10}$)}\tabularnewline
\hline 
$1B_{3u}^{-}$&
3.25&
2.91&
\multicolumn{4}{c|}{3.47\cite{platt-naphtha}, 3.57\cite{biermann}, 3.72\cite{dick83}}&
\multicolumn{8}{c|}{3.72\cite{pariser24}, 3.22\cite{ham}, 3.23\cite{rama-anthra,kawashima102},
3.84\cite{rubio}, 3.93\cite{rubio}$^{*}$, 3.60\cite{heinze113-hepta},
3.89\cite{grimme}}\tabularnewline
\hline 
$1B_{2u}^{+}$&
3.66&
3.55&
\multicolumn{4}{c|}{3.31\cite{platt-naphtha}, 3.38\cite{biermann}, 3.42\cite{dick83,man-trajmar25},
3.43\cite{lambert81}}&
\multicolumn{8}{c|}{3.65\cite{pariser24,hummer}, 3.60\cite{ham}, 3.68\cite{rama-anthra},
3.40\cite{kawashima102}, 3.21\cite{rubio}, 4.05\cite{rubio}$^{*}$,
2.96\cite{heinze113-hepta}, 3.69\cite{grimme}}\tabularnewline
\hline 
$1B_{3u}^{+}$&
5.34&
4.64&
\multicolumn{4}{c|}{4.83\cite{platt-naphtha}, 4.86\cite{biermann}, 5.24\cite{man-trajmar25,lyons4}}&
\multicolumn{8}{c|}{5.50\cite{pariser24}, 5.27\cite{ham}, 5.36\cite{rama-anthra}, 4.77\cite{kawashima102},
5.35\cite{hummer}, 5.14\cite{rubio}, 5.49\cite{rubio}$^{*}$, 4.93\cite{heinze113-hepta}}\tabularnewline
\hline 
\multicolumn{15}{|c|}{Tetracene (C$_{\text{18}}$H$_{12}$)}\tabularnewline
\hline 
$1B_{3u}^{-}$&
3.22&
3.02&
\multicolumn{4}{c|}{3.16\cite{platt-naphtha}, 3.32\cite{biermann}, 3.12\cite{berlman-book}}&
\multicolumn{8}{c|}{3.57\cite{pariser24}, 3.04\cite{ham}, 2.92\cite{kawashima102},
3.46\cite{rubio}, 3.58\cite{rubio}$^{*}$, 3.21\cite{heinze113-hepta},
3.52\cite{grimme}}\tabularnewline
\hline 
$1B_{2u}^{+}$&
3.16&
2.97&
\multicolumn{4}{c|}{2.60\cite{platt-naphtha,bree-lyons}, 2.71\cite{biermann}, 2.72\cite{berlman-book},
2.63\cite{birk-book}}&
\multicolumn{8}{c|}{3.11\cite{pariser24}, 3.05\cite{ham}, 2.80\cite{kawashima102},
2.44\cite{rubio}, 3.22\cite{rubio}$^{*}$, 2.19\cite{heinze113-hepta},
2.90\cite{grimme}}\tabularnewline
\hline 
$1B_{3u}^{+}$&
5.01&
4.38&
\multicolumn{4}{c|}{4.55\cite{platt-naphtha}, 4.52\cite{biermann}, 4.50\cite{bree-lyons},
4.51\cite{birk-book}}&
\multicolumn{8}{c|}{5.09\cite{pariser24}, 4.86\cite{ham}, 4.32\cite{kawashima102},
4.63\cite{rubio}, 4.96\cite{rubio}$^{*}$, 4.38\cite{heinze113-hepta}}\tabularnewline
\hline 
\multicolumn{15}{|c|}{Pentacene (C$_{\text{22}}$H$_{\text{14}}$)}\tabularnewline
\hline 
$1B_{3u}^{-}$&
3.17&
2.99&
\multicolumn{4}{c|}{2.96\cite{platt-naphtha}, 3.05\cite{biermann}, 2.89\cite{birk-book},
3.73\cite{penta-3}}&
\multicolumn{8}{c|}{3.51\cite{pariser24}, 2.99\cite{ham}, 3.20\cite{rubio}, 3.34\cite{rubio}$^{*}$,
2.95\cite{heinze113-hepta}, 3.12\cite{grimme}}\tabularnewline
\hline 
$1B_{2u}^{+}$&
2.86&
2.65&
\multicolumn{4}{c|}{2.14\cite{platt-naphtha}, 2.23\cite{biermann}, 2.12\cite{birk-book},
2.28\cite{penta-3}}&
\multicolumn{8}{c|}{2.81\cite{pariser24}, 2.70\cite{ham}, 1.90\cite{rubio}, 2.64\cite{rubio}$^{*}$,
1.65\cite{heinze113-hepta}, 2.37\cite{grimme}}\tabularnewline
\hline 
$1B_{3u}^{+}$&
4.73&
4.22&
\multicolumn{4}{c|}{4.01\cite{platt-naphtha}, 4.10\cite{birk-book}, 4.40\cite{penta-3}}&
\multicolumn{8}{c|}{4.80\cite{pariser24}, 4.57\cite{ham}, 4.24\cite{rubio}, 4.57\cite{rubio}$^{*}$,
3.96\cite{heinze113-hepta}}\tabularnewline
\hline 
\multicolumn{15}{|c|}{Hexacene (C$_{26}$H$_{\text{16}}$)}\tabularnewline
\hline 
$1B_{3u}^{-}$&
3.07&
2.77&
\multicolumn{4}{c|}{2.80\cite{biermann}, 2.67\cite{hexa-1}}&
\multicolumn{8}{c|}{3.34\cite{hexa-1}, 3.01\cite{rubio}, 3.17\cite{rubio}$^{*}$, 2.76\cite{heinze113-hepta},
2.87\cite{grimme}}\tabularnewline
\hline 
$1B_{2u}^{+}$&
2.71&
2.38&
\multicolumn{4}{c|}{1.90\cite{biermann}, 1.91\cite{hexa-1}}&
\multicolumn{8}{c|}{2.18\cite{hexa-1}, 1.50\cite{rubio}, 2.23\cite{rubio}$^{*}$, 1.25\cite{heinze113-hepta},
2.02\cite{grimme}}\tabularnewline
\hline 
$1B_{3u}^{+}$&
4.61&
4.07&
\multicolumn{4}{c|}{3.94\cite{hexa-1}}&
\multicolumn{8}{c|}{4.05\cite{hexa-1}, 3.94\cite{rubio}, 4.27\cite{rubio}$^{*}$, 3.64\cite{heinze113-hepta}}\tabularnewline
\hline 
\multicolumn{15}{|c|}{Heptacene (C$_{\text{30}}$H$_{1\text{8}}$)}\tabularnewline
\hline 
$1B_{3u}^{-}$&
2.73&
2.35&
\multicolumn{4}{c|}{-}&
\multicolumn{8}{c|}{2.35\cite{heinze113-hepta}}\tabularnewline
\hline 
$1B_{2u}^{+}$&
2.63&
2.24&
\multicolumn{4}{c|}{-}&
\multicolumn{8}{c|}{0.94\cite{heinze113-hepta}}\tabularnewline
\hline 
$1B_{3u}^{+}$&
4.48&
3.80&
\multicolumn{4}{c|}{-}&
\multicolumn{8}{c|}{3.36\cite{heinze113-hepta}}\tabularnewline
\hline
\end{tabular}
\par\end{raggedright}
\end{table}

Before discussing the individual oligomers, we would like to summarize
the patterns which emerge from the computed optical absorption spectra
of acenes ranging from naphthalene to heptacene. The absorption spectrum
of all the oligoacenes, irrespective of the P-P-P parameters used
in the calculations, contains following important features:

\begin{enumerate}
\item The first peak in the absorption spectrum for all the oligomers studied
corresponds to the $1B_{2u}^{+}$ excited state of the system. The
most important configuration contributing to the many-particle wave
function of the state corresponds to the $|H\rightarrow L\rangle$
excitation, in agreement with H\"uckel model calculations. The relative
intensity of this feature decreases with the increasing size of the
oligomer, again in agreement with the H\"uckel model results.
\item The second, and the most intense feature in the spectrum corresponds
to the $1B_{3u}^{+}$ state, obtained by the absorption of a $x$-polarized
photon. The most important configurations contributing to the wave
function of this state, for $n=\mbox{even}$ oligomers are excitations
$|H\rightarrow L+n/2\rangle$ and $|H-n/2\rightarrow L\rangle$, and
for $n=\mbox{odd}$, the excitations $|H\rightarrow L+(n\pm1)/2\rangle$
and $|H-(n\pm1)/2\rightarrow L\rangle$. This aspect of the correlated
spectrum is also in good agreement with the H\"uckel model results.
\item Another important state, namely $1B_{3u}^{-}$ state exists for all
oligoacenes. Because it has the same particle-hole symmetry ($-$)
as the ground state, in P-P-P (and H\"uckel) calculations it does
not contribute to the absorption spectrum. But many experiments report
this state as a very weak feature in the absorption spectrum. In our
calculations, for naphthalene and anthracene, this state occurs at
energies lower than the $1B_{2u}^{+}$ state, but for longer oligomers,
it is at higher excitation energies than the $1B_{2u}^{+}$ state.
The important configurations contributing to the wave function of
this state are the same as the ones contributing to $1B_{3u}^{+}$,
except for their opposite relative signs, for all the oligomers up
to pentacene. But, from hexacene onwards, it is the doubly excited
configurations which contribute significantly to this state. Thus,
it is the electron-correlation effects which are responsible for its
distinct location in the spectrum as compared to the $1B_{3u}^{+}$
state. 
\item Generally, the spectra computed with the standard parameters are in
good qualitative agreement with those computed with the screened parameters.
Quantitatively speaking, screened parameter spectra are redshifted
as compared to the standard parameter ones, and are in better agreement
with the experiments for longer acenes.
\item Although there are some qualitative similarities between the P-P-P
and H\"uckel model results, quantitatively speaking the optical gaps
obtained using the H\"uckel model are much smaller than their P-P-P
and experimental counterparts.
\end{enumerate}
Having emphasized the general features of our work, next we discuss
the results of our calculations for individual oligomers, and compare
them with the theoretical results of other authors, and the experimental
ones. Quantitative aspects of our calculations are summarized in table
\ref{tab:comparion}, which also reports some of the experimental
and the theoretical results of other authors. Additionally, in Appendix
\ref{sec:appa}, we present detailed tables for individual oligomers
containing the excitation energies and many-particle wave functions
of important excited states, as well as corresponding transition dipoles.

\subsubsection{Naphthalene}

\begin{figure}[H]
\begin{centering}
\includegraphics[width=10cm,keepaspectratio]{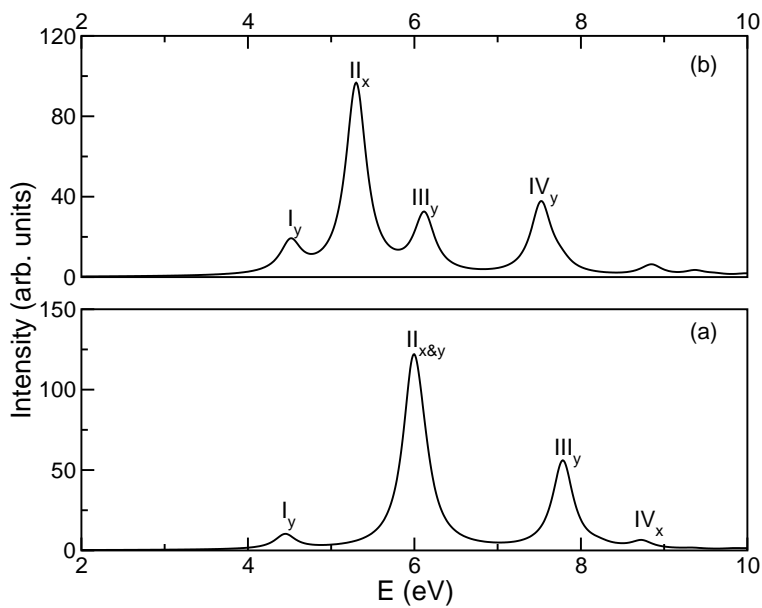}
\par\end{centering}

\caption{Linear optical absorption spectrum of naphthalene computed with (a)
standard parameters and (b) screened parameters in the P-P-P model
Hamiltonian. FCI method was used for the purpose. A line width of
0.15 eV was assumed. As a subscript of each feature, its polarization
is also mentioned. Thus, \emph{e.g.}, III$_{y}$ implies that peak
III is short-axis polarized.}

\label{fig-acene2} 
\end{figure}

In Fig.\ref{fig-acene2} we present the linear optical absorption
spectra of naphthalene computed with the FCI method, using both the
standard and the screened parameters in the P-P-P Hamiltonian. The
wave functions of the excited states contributing to various peaks
are presented in Tables \ref{tab-acene2-fci} and \ref{acene2-scr-fci-tab}
in the appendix. Since these calculations were performed using the
FCI approach, they are exact within the model chosen, and cannot be
improved. Therefore, any discrepancy which these results may exhibit
with respect to the experiments, is a reflection of the limitations
of the model or the parameters used, and not that of the correlation
approach.

From Fig.\ref{fig-acene2} and Tables \ref{tab-acene2-fci} and \ref{acene2-scr-fci-tab},
it is obvious that the first peak of the spectra calculated with the
standard and screened parameters are in excellent qualitative and
quantitative agreement with each other. The second peak computed using
the standard parameters is due to both the $x$-polarized and the
$y$-polarized components, but it is obvious from the Table \ref{tab-acene2-fci}
that $x$-polarized component dominates. However, with the screened
parameters, it gets dissociated into two separate features II and
III with $x$-polarized and $y$-polarized components, respectively.
Qualitatively, both the spectra are in good agreement with each other
(see Tables \ref{tab-acene2-fci} and \ref{acene2-scr-fci-tab}).
Regarding the quantitative aspects, one notices that the excitation
energies computed with the screened parameters are red shifted as
compared to those computed with the standard parameters. This is in
agreement with results obtained for other conjugated polymers as well\cite{shukla-ppv,shukla-tpa,shukla-thg,sony-pdpa}.

Comparison of our results for naphthalene, for the invisible (dipole
forbidden) $1B_{3u}^{-}$ state and the first two visible features,
$1B_{2u}^{+}$, and $1B_{3u}^{+}$ states with the experimental ones,
and those of other calculations, is presented in Table \ref{tab:comparion}.
Although, several other experimental results on optical absorption
in naphthalene exist,\cite{platt-naphtha,birks-naphtha,berg-naphtha,aleksandrovsky,huebner-naphtha,naphth-2,dick-holhneicher,biermann}
but here we compare our results with the experimental results of Huebner
\emph{et al}.\cite{huebner-naphtha} and Aleksandrovsky \emph{et al}.,\cite{aleksandrovsky}
who performed experiments on gas-phase of naphthalene. Since we have
performed calculations on isolated oligomers, therefore, most appropriate
comparison of our results will be with the data obtained by gas-phase
experiments. Huebner \emph{et al}.\cite{huebner-naphtha} suggested
$1B_{2u}^{+}$ and $1B_{3u}^{+}$ states at 4.45 eV and 5.89 eV respectively,
while Aleksandrovsky \emph{et al}.\cite{aleksandrovsky} mentioned
these states to be at 4.46 eV and 5.95 eV respectively. Thus, our
results on the relative ordering of the $1B_{2u}^{+}$ and the $1B_{3u}^{+}$
states in the spectra, using standard parameters are in excellent
agreement with the results of Huebner \emph{et al}.\cite{huebner-naphtha}
and the recent experiment of Aleksandrovsky \emph{et al}.\cite{aleksandrovsky}
The screened parameters slightly overestimate the experimental results\cite{aleksandrovsky,huebner-naphtha}
for $1B_{2u}^{+}$ state and underestimate them for $1B_{3u}^{+}$
state. Excellent agreement between our results and those of recent
experimental results of Aleksandrovsky \emph{et al}.\cite{aleksandrovsky}
testifies to the essential correctness of the P-P-P model to the naphthalene
with standard parameters. The dipole forbidden $1B_{3u}^{-}$ state
has not been discussed by Aleksandrovsky \emph{et al}. but Huebner
\emph{et al}. suggested this feature to be at 4.0 eV which is higher
as compared to both of our results. Thus, both sets of calculations
appear to capture the qualitative features of the spectra quite well,
however, quantitatively, standard parameter calculations appear to
be more accurate upon comparison with the experiments.\cite{aleksandrovsky,huebner-naphtha}

Several theoreticians have studied the low-lying excited states of
naphthalene.\cite{pariser24,ham,tavan,simmons-naphtha,ramasesha-soos91,nakatsuji142,hashimoto104,hofer-naphta,lipari-duke63,baldo77,heinze113-hepta,hummer-prb71,grimme,rubio}
The results obtained from our standard set of parameters for the $1B_{2u}^{+}$
and the $1B_{3u}^{+}$ states are in perfect agreement with the calculations
performed by Pariser \emph{et al}.\cite{pariser24} and Ramasesha
\emph{et al}.\cite{ramasesha-soos91} Also, they agree quite well
with the recent calculations performed by Rubio and coworkers,\cite{rubio}
using B3LYP as exchange-correlation (xc) functional. The dipole forbidden
$1B_{3u}^{-}$ state is obtained at lower energy than calculated by
Pariser \emph{et al}.\cite{pariser24} and Rubio and coworkers,\cite{rubio}
but is in excellent agreement with the results of Ramasesha \emph{et
al}.\cite{ramasesha-soos91}

\subsubsection{Anthracene}

An anthracene molecule contains 14 $\pi$-electrons, which is quite
a large system for the application of the FCI approach. However, the
FCI calculations using the P-P-P model have been performed by Ramasesha
\emph{et al}.\cite{rama-anthra} for anthracene, but they did not
report its linear absorption spectrum. Therefore, in our present work
we have computed the linear absorption spectrum of the anthracene
using the FCI approach, and have also discussed the nature of low-lying
excited states contributing to the spectrum.

\begin{figure}[H]
\begin{centering}
\includegraphics[width=10cm,keepaspectratio]{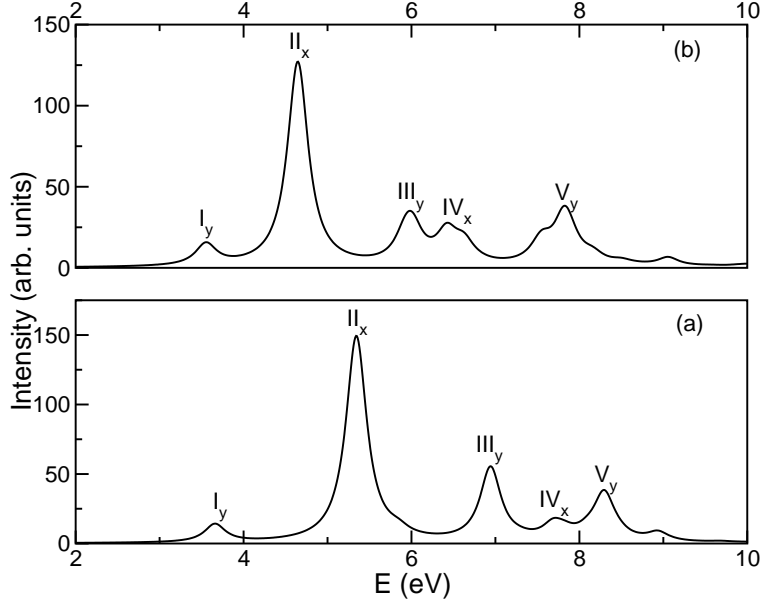}
\par\end{centering}

\caption{Linear optical absorption spectrum of anthracene computed with (a)
standard parameters and (b) screened parameters. FCI method coupled
with the P-P-P model Hamiltonian was used for the purpose. A line
width of 0.15 eV was assumed. As a subscript of each feature, its
polarization is also mentioned.}

\label{Fig:acene3} 
\end{figure}

Fig.\ref{Fig:acene3} presents the linear optical absorption spectra
of anthracene computed using the FCI method, along with the standard
and screened sets of parameters in the P-P-P Hamiltonian. The wave
functions of the excited states contributing to various peaks are
presented in Tables \ref{acene3-fci-tab} and \ref{acene3-scr-fci-tab}
in the appendix. 

In our calculations on anthracene using standard parameters, the first
peak of the spectrum was identified as $1B_{2u}^{+}$ state at 3.66
eV, while the most intense feature of the spectrum corresponding to
the $1B_{3u}^{+}$ state, was obtained at 5.34 eV. The dipole forbidden
$1B_{3u}^{-}$ state was obtained at 3.25 eV and its wave function
contains both singly and doubly excited configurations. Using screened
parameters, we obtained $1B_{3u}^{-}$, $1B_{2u}^{+}$, and $1B_{3u}^{+}$
states at 2.91eV, 3.55 eV, and 4.64 eV, respectively. Thus, calculations
performed with both sets of P-P-P parameters predict the dipole forbidden
$1B_{3u}^{-}$ state at lower energies than the $1B_{2u}^{+}$ state. 

Comparison of our theoretical results, to the experimental and theoretical
results of other authors, is presented in Table \ref{tab:comparion}.
Several experimental investigations of the absorption spectra of anthracene
have been performed over the years. For the $1B_{2u}^{+}$ state,
our results with both sets of parameters overestimate the experimental
results of Klevens and Platt (3.31 eV),\cite{platt-naphtha}, Biermann
\emph{et al}. (3.38 eV),\cite{biermann} Lambert \emph{et al}. (3.43
eV),\cite{lambert81} Dick \emph{et al}. (3.42 eV),\cite{dick83}
and Man \emph{et al}. (3.424 eV).\cite{man-trajmar25} Comparatively
speaking, however, the result obtained from screened parameters is
in slightly better agreement with the experimental results.

Lambert \emph{et al}. \cite{lambert81} studied the isolated molecules
of anthracene using jet spectroscopy, while Dick \emph{et al}. \cite{dick83}
observed the low-lying excited states of anthracene in gas-phase.
Thus, our calculations, which were performed on isolated oligomers,
are directly comparable to these two experimental results. The most
intense feature of the spectra, which corresponds to $1B_{3u}^{+}$
state was obtained at 5.34 eV from standard parameters while screened
parameters suggested this state to be at 4.64 eV. Our result with
standard parameters agrees well with the results obtained by Man \emph{et
al}.,\cite{man-trajmar25} and Lyons \emph{et al}.\cite{lyons4} They
both suggested this state to be at 5.24 eV. Our screened parameter
result is in good agreement with the result of Klevens and Platt,\cite{platt-naphtha}
and Biermann \emph{et al}.\cite{biermann} for the $1B_{3u}^{+}$
state, who suggested this state to be at 4.83 eV and 4.86 eV, respectively.

Our correlated value of 3.25 eV for the excitation energy of the $1B_{3u}^{-}$
state obtained using the standard parameters is in fairly good agreement
with the experimental results of Klevens and Platt (3.47 eV)\cite{platt-naphtha}
and Biermann \emph{et al}. (3.57 eV).\cite{biermann} However, it
is lower as compared to the experimental results of Dick \emph{et
al}. (3.72 eV).\cite{dick83} Our screened parameter results (2.91
eV) obviously significantly underestimate the experimental excitation
energy of the $1B_{3u}^{-}$ state. Thus, we observe that for some
states our standard parameter results agree well with the experiments,
while for other states screened parameter results are in better agreement
with the experiments. However, the standard-parameter-based results
are in better overall agreement with the gas-phase experiments, as
compared to those obtained using screened parameters.

Next we discuss and compare our results with the results of calculations
performed by other theoreticians. Our correlated results obtained
by using standard parameters for $1B_{3u}^{-}$, $1B_{2u}^{+}$, and
$1B_{3u}^{+}$ states are in perfect agreement with the results obtained
by Ramasesha \emph{et al}.,\cite{rama-anthra} using FCI methodology
with P-P-P model Hamiltonian. Again, this perfect agreement proves
the essential correctness of our calculations. The $1B_{2u}^{+}$
state obtained using standard parameters shows an excellent agreement
with the pioneering work of Pariser,\cite{pariser24} who performed
the SCI calculations. However, energetically both $1B_{3u}^{+}$ (dipole
allowed) and $1B_{3u}^{-}$ (dipole forbidden) states obtained from
our calculations are lower than those reported by Pariser.\cite{pariser24}
This clearly is due to the superior treatment of electron correlations
in our work. Additionally, the results for the $1B_{2u}^{+}$, and
$1B_{3u}^{+}$ states are in excellent agreement with results of Hummer
\textit{et al.}\cite{hummer} and in close agreement with the recent
results of Rubio and coworkers,\cite{rubio} obtained using B3LYP
as xc functional. Recently, Kawashima and co-workers\cite{kawashima102}
performed calculations on linear absorption in anthracene using \emph{ab
initio} MRMP methodology. The energies they obtained for $1B_{2u}^{+}$
state (3.40 eV) and $1B_{3u}^{+}$ state (4.77 eV) are quite close
to our screened parameter results, and slightly less than the results
obtained from our calculations using standard parameters. They mentioned
$1B_{3u}^{-}$ state at 3.23 eV which is in perfect agreement with
the results which we have obtained using standard parameters, while
it is higher than the value which we have obtained using screened
parameters (2.91 eV). Few other theoretical results of other authors
are also given in the Table \ref{tab:comparion} for the sake of comparison.

\subsubsection{Tetracene}

\begin{figure}[H]
\begin{centering}
\includegraphics[width=11cm,keepaspectratio]{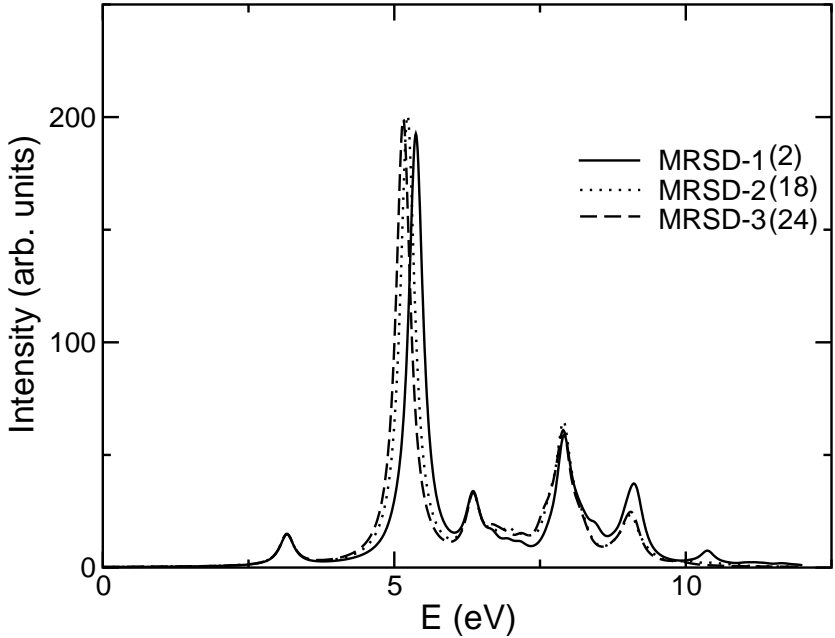}
\par\end{centering}

\caption{Convergence of linear absorption spectrum of tetracene, computed
using the MRSDCI method (standard parameters) with respect to number
of reference configurations ($N_{ref}$). The numbers written in the
parenthesis represent $N_{ref}$ included in the corresponding MRSDCI
calculations.\label{fig:convergence-of-mrsd}}
\end{figure}

The next member of the polyacene family is tetracene, which contains
18 $\pi$-electrons. Because of relatively larger number of $\pi$-electrons
in the system, it is virtually impossible to use the FCI method for
the present system. Thus, we used QCI technique for computing the
ground state and the low-lying excited states corresponding to the
short-axis polarized transitions ($B_{2u}$-type states), while MRSDCI
methodology for the excited states which correspond to the long-axis
polarized transitions($B_{3u}$-type states). The reason behind using
different CI techniques is that the QCI method can only be used for
single reference states such as the ground state ($1A_{g}$) and the
$B_{2u}$ type states, while from previous Tables \ref{tab-acene2-fci},
\ref{acene2-scr-fci-tab}, \ref{acene3-fci-tab}, and \ref{acene3-scr-fci-tab},
it is obvious that $B_{3u}$ states are multi-reference states with
two dominant configurations for which QCI method cannot be used. Therefore,
for states of $B_{3u}$ symmetry, the MRSDCI method has been used.

\begin{figure}[H]
\begin{centering}
\includegraphics[width=10cm,keepaspectratio]{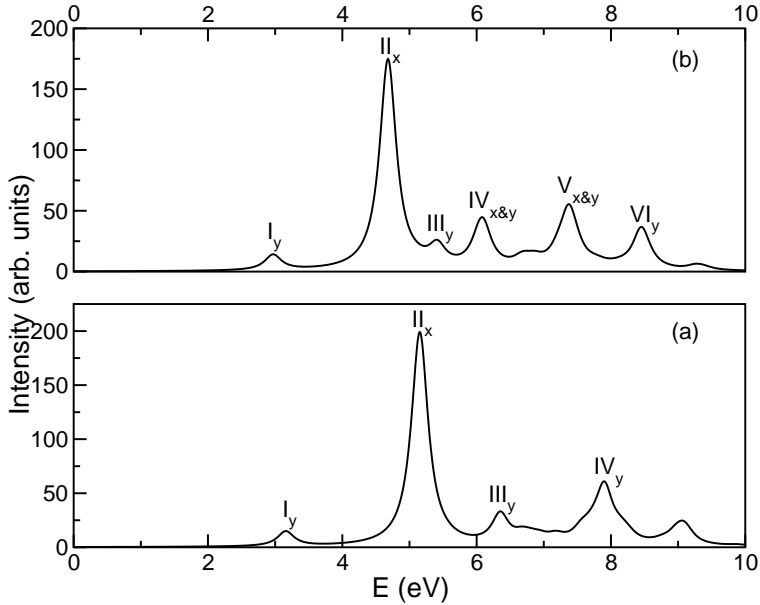}
\par\end{centering}

\caption{Linear optical absorption spectrum of tetracene computed with (a)
standard parameters, and (b) screened parameters. QCI method was used
to compute the ground state ($1A_{g}^{-}$ state) and $B_{2u}^{+}$
states, while MRSDCI method was used to compute $B_{3u}^{+}$ states.}

\label{fig:acene4} 
\end{figure}

In order to demonstrate the convergence of our MRSDCI calculations,
we present the spectra in Fig.\ref{fig:convergence-of-mrsd}, calculated
with increasing number of reference configurations ($N_{ref}$). It
is obvious from the figure that the spectrum computed using twenty
four reference configurations is in very good agreement with the spectrum
computed using eighteen reference configurations, implying that the
convergence with respect to $N_{ref}$ has been achieved. We would
also like to mention that the accuracy of the MRSDCI method was demonstrated
in our earlier works where the results obtained using that approach
were found to be in excellent agreement with the experiments.\cite{shukla2,shukla-ppv,shukla-ppp}

Tetracene has been studied by several theoreticians\cite{ham,pariser24,tavan,kawashima102,bredas18,lipari-duke63,baldo77,grimme,heinze113-hepta,houk-hexa,hummer-prb71,rubio,Ramasesha-2}
as well as experimentalists.\cite{platt-naphtha,sebastian61,berlman-book,bree-lyons,birk-book,burrow86,dahlberg71,biermann}
The linear absorption spectra of tetracene, computed using standard
and screened parameters are presented in Fig.\ref{fig:acene4}. The
excited states together with the corresponding energies, transition
dipoles, and wave functions are presented in Tables \ref{acene4-mrsd3-qci}
and \ref{acene4-scr-mrsd3-qci}. 

The first peak of the correlated spectra was obtained at 3.16 eV and
2.97 eV, using the standard parameters and the screened parameters,
respectively, and corresponds to the $1B_{2u}^{+}$ state of the oligomer.
The second peak, which is also the most intense feature in both the
spectra, corresponds to the $1B_{3u}^{+}$ state. The dipole forbidden
$1B_{3u}^{-}$ state has been obtained at higher energy than the \textbf{$1B_{2u}^{+}$}
state (\emph{cf.} Tables \ref{acene4-mrsd3-qci} and \ref{acene4-scr-mrsd3-qci}),
which is different as compared to naphthalene and anthracene, but
in agreement with the results of other investigators (\emph{cf.} Table
\ref{tab:comparion}).

The comparison of our calculated results for $1B_{3u}^{-}$, $1B_{2u}^{+}$,
and $1B_{3u}^{+}$ excited states with the several experimental and
other calculated results is present in Table \ref{tab:comparion}.
Our standard parameter results generally overestimate the excitation
energies when compared to the experiments, while the agreement between
the screened parameters results and the experiments\cite{platt-naphtha,berlman-book,bree-lyons,birk-book,biermann}
is much better.

The $1B_{2u}^{+}$ excitation energies computed using both the standard
as well as the screened parameters overestimate all the experimental
results.\cite{platt-naphtha,berlman-book,bree-lyons,birk-book,sebastian61,biermann}
Yet our screened parameter value of the excitation energy of $1B_{2u}^{+}$
state (2.97 eV) is in reasonably good agreement with the experimental
values of Klevens and Platt (2.60 eV),\cite{platt-naphtha} Biermann
\textit{et al.} (2.71 eV),\cite{biermann} Bree and Lyons (2.60 eV),\cite{bree-lyons}
Birks (2.63 eV),\cite{birk-book} and Berlman (2.72 eV).\cite{berlman-book}
As far as the $1B_{3u}^{+}$ state is concerned, our screened parameter
value of 4.38 eV is fairly close to the several experimental values
reported to be near 4.50 eV. \cite{platt-naphtha,bree-lyons,birk-book,biermann}
However, our standard parameter value of 5.01 eV for the same overestimates
the experiments by about 0.5 eV. Our results for the dipole forbidden
$1B_{3u}^{-}$ state with both sets of parameters are in good agreement
with the experiments. As compared to the experiments of Klevens and
Platt\cite{platt-naphtha} and Berlman\cite{berlman-book}, our standard
parameter excitation energy for the $1B_{3u}^{-}$ is slightly higher,
while the screened parameter value is slightly lower. Thus, we can
conclude that in case of tetracene, for all the three excited states
discussed, screened parameter based calculations provide better agreement
with the experiments.

On comparing our results with other theoretical results we find that
$1B_{2u}^{+}$ and $1B_{3u}^{+}$ states, computed using standard
parameters are in excellent agreement with the benchmark work of Pariser,\cite{pariser24}
while the $1B_{3u}^{-}$ state has been computed at lower value by
both of our parameters. Ramasesha and co-workers also computed the
optical gap of tetracene using P-P-P model Hamiltonian coupled with
the DMRG technique.\cite{Ramasesha-2} They reported the $1B_{2u}^{+}$
state at 3.20 eV, which is in perfect agreement with the $1B_{2u}^{+}$
state computed by us using the standard parameters. The results obtained
using screened parameters for all the compared states are in very
good agreement with the results of Ham \emph{et al}.\cite{ham} and
with the \emph{ab initio} MRMP based results of Kawashima \emph{et
al.}\cite{kawashima102} The results of Rubio and coworkers\cite{rubio}
obtained using B3LYP as xc functional for the $x$-polarized states
$1B_{3u}^{-}$ and $1B_{3u}^{+}$ are higher than our results of the
screened parameters, while is at low energy for the $y$-polarized
$1B_{2u}^{+}$ state.

\subsubsection{Pentacene}

The most thoroughly studied and most widely used member of the polyacene
family is pentacene. It has been quite famous in recent years due
to its use in thin film growth of electronic devices.\cite{street-apl80,klauk46,klauk20}
It is among the most promising organic molecular semiconductor due
to its high charge carrier mobility.\cite{gundlach18,nelson-apl72,butko-apl83}
It has been studied by several researchers to understand the nature
of its low-lying excited states. However, most of the experimentalists
have mainly reported the energy corresponding to  the $1B_{2u}^{_{+}}$
state. \cite{dahlberg71,sebastian61,penta-2,park-apl80,kim-tsf420,puigdollers-tsf427,he-apl84,lee-apl84}
As per our knowledge, experimentally only Klevens and Platt,\cite{platt-naphtha}
Biermann \textit{et al.},\cite{biermann} and Birks\cite{birk-book}
have performed extensive studies of the $1B_{3u}^{-}$, $1B_{2u}^{_{+}}$,
and $1B_{3u}^{+}$ excited states. Additionally, Halasinski \emph{et
al.}\cite{penta-3} measured the vibronic transitions of neutral pentacene
isolated in Ne, Ar, and Kr matrices and reported the data corresponding
to the $1B_{3u}^{-}$, $1B_{2u}^{_{+}}$, and $1B_{3u}^{+}$ excited
states.

\begin{figure}[H]
\begin{centering}
\includegraphics[width=10cm,keepaspectratio]{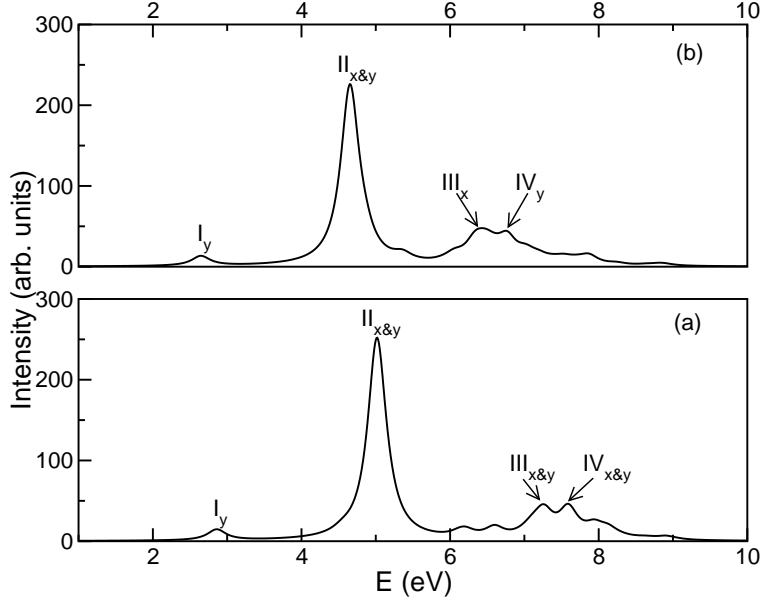}
\par\end{centering}

\caption{Linear optical absorption spectrum of pentacene computed with (a)
standard parameters and (b) screened parameters. QCI method was used
to compute the $B_{2u}^{+}$ states, while MRSDCI method was used
to compute the $B_{3u}^{+}$ states.}

\label{fig:acene5} 
\end{figure}

Theoretically, the low-lying states of pentacene have been studied
by several people, using various methodologies.\cite{ham,pariser24,lipari-duke63,baldo77,tiago-prb67,hummer-prb71,Ramasesha-2}
Similar to tetracene, for pentacene also we used the QCI method for
computing the ground state ($1A_{g}$) and the $B_{2u}$-type excited
states, while the $B_{3u}$-type excited states were computed using
the MRSDCI technique.

In Fig.\ref{fig:acene5} we present the linear optical absorption
spectra computed using both standard and screened parameters in the
P-P-P model Hamiltonian. The energies and wave functions corresponding
to the visible features in the spectra as well as $1B_{3u}^{-}$ state,
are presented in the Tables \ref{acene5-mrsd3-qci-tab} and \ref{acene5-scr-mrsd3-qci}
in the appendix. 

The first peak as usual, corresponds mainly to the $1B_{2u}^{+}$
state, while the second peak is a mixture of states corresponding
to $x$-polarized and $y$-polarized photons. But, as is clear from
the transition dipoles, the intensity of the peak is mainly due to
$x$-polarized photon corresponding to $H\rightarrow L+2+c.c.$ excitations
leading to the $1B_{3u}^{+}$ state (see Tables \ref{acene5-mrsd3-qci-tab}
and \ref{acene5-scr-mrsd3-qci}). 

The comparison of our results with other theoretical and experimental
results is presented in the Table \ref{tab:comparion}. On comparing
our results with the experimental results we found that on most of
the occasions both sets of parameters overestimate the experimental
results. However, overall, the results obtained using the screened
parameters are in much better agreement with the experiments. For
example, the screened parameter results on $1B_{2u}^{+}$ and $1B_{3u}^{+}$
states are in very good agreement with the experiments of Klevens
and Platt,\cite{platt-naphtha} Biermann \textit{et al.},\cite{biermann}
Birks,\cite{birk-book} and Halasinski \emph{et al.}\cite{penta-3}
However, several experimentalists have reported that the $1B_{2u}^{+}$
state lies between 1.7 eV to 1.9 eV.\cite{park-apl80,kim-tsf420,puigdollers-tsf427,he-apl84,lee-apl84}
The difference in our calculated results and these experimental results
could be due to the bulk effects, as most of these experiments were
performed on bulk pentacene, thus possibly explaining substantially
lower values of band gaps. We again emphasize that our calculations
were performed on single molecule of pentacene, leading to comparatively
larger excitation energies.

Next, we compare our results with the theoretical results of other
authors. Our correlated results for $1B_{2u}^{+}$ and $1B_{3u}^{+}$
obtained using standard parameters are in very good agreement with
Pariser\cite{pariser24} while the energy corresponding to $1B_{3u}^{-}$
state is obtained to be lower in our standard parameter calculations.
Our standard parameter value of $1B_{2u}^{+}$ excitation energy (2.86
eV) is also in excellent agreement with the value 2.92 eV obtained
by Raghu \emph{et al.\cite{Ramasesha-2}} from their DMRG calculations.
Our results obtained using screened parameters are in very good agreement
with the results obtained by Ham \emph{et al}.\cite{ham} They used
free electron molecular orbitals (FE MO) theory to compute the energies
of the low-lying excited states. From our screened parameter calculations,
the energy obtained for the $1B_{3u}^{-}$ excited state is in perfect
agreement with the results of Ham \emph{et al}.,\cite{ham} while
the energies of $1B_{2u}^{+}$ and $1B_{3u}^{+}$ excited states are
slightly less than that of Ham \emph{et al}.\cite{ham} The energies
of the $x$-polarized $1B_{3u}^{-}$ and $1B_{3u}^{+}$ states obtained
using standard as well as screened parameters agree well with the
results for these states calculated by Rubio and coworkers,\cite{rubio}
using xc functional. Recently, Tiago \emph{et al}.\cite{tiago-prb67}
computed the energy gap for solution-phase crystallized (S) structure
and vapor-phase crystallized (V) structure of pentacene using \emph{ab
initio} pseudo-potential density functional method and GW approach.
They obtained the energy gap to be 2.2 eV for S structure while 1.9
eV for V structure. This low value of energy can be understood due
to the bulk effects. Thus, their calculation supports our belief that
bulk effects affect the $H\rightarrow L$ gap significantly.

\subsubsection{Hexacene}

With the increase in the size of oligoacene, it becomes less stable,
poorly soluble and more reactive.\cite{clar&many-hexa,aihara-hexa,sauer-hexa,bachmann-hexa,herndon-hexa,wang-hexa}
Thus, preparation and practical study of higher acenes like hexacene
is a difficult task\cite{campbell-hexa,clar-hexa6}. To the best of
our knowledge, experimentally hexacene has only been studied by Biermann\cite{biermann}
and Angliker \emph{et al}.\cite{hexa-1} They measured the absorption
spectra of hexacene in solution. Angliker \emph{et al}.\cite{hexa-1}
have also computed the low-lying excited states by performing P-P-P
SCF-SCI calculations. Most recently, Grimme \textit{et al.},\cite{grimme}
Heinze \emph{et al}.,\cite{heinze113-hepta} and Houk \emph{et al}.\cite{houk-hexa}
have also calculated the low-lying excited states using the time-dependent
density functional theory (TDDFT) based approaches.

\begin{figure}[H]
\begin{centering}
\includegraphics[width=10cm,keepaspectratio]{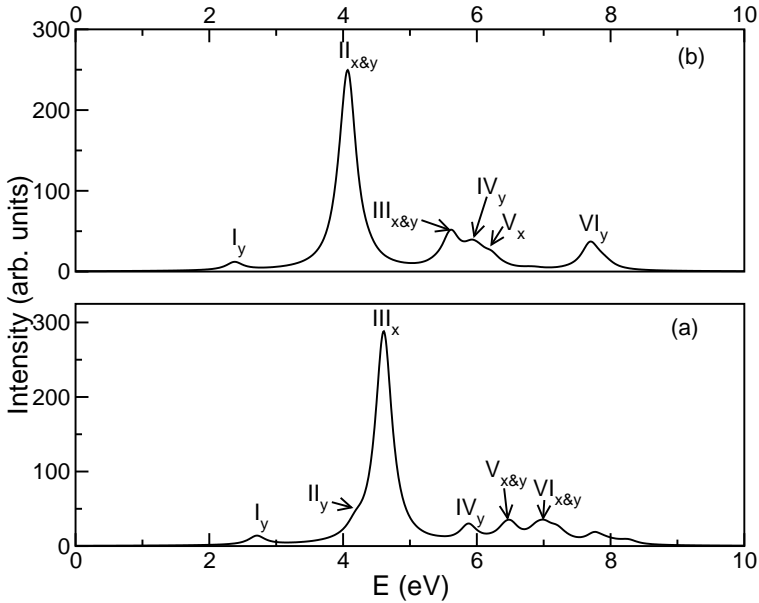}
\par\end{centering}

\caption{Linear optical absorption spectrum of hexacene computed with (a)
standard parameters and (b) screened parameters. MRSDCI method was
used to include the electron correlation effects.}

\label{fig:acene6} 
\end{figure}

In Fig.\ref{fig:acene6}, we present the linear absorption spectra
of hexacene, computed using both standard and screened parameters
in the P-P-P model Hamiltonian. For inclusion of electron correlation
effects, the MRSDCI method was employed for all the states involved.
The excited states, corresponding energies, transition dipoles, and
wave functions of the various features of the spectra are presented
in the Tables \ref{acene6-mrsd4-tab} and \ref{acene6-scr-mrsd4}
in the appendix. Spectrum obtained using standard parameters predicts
the $1B_{2u}^{+}$ state at 2.71 eV, while from screened parameters
it was obtained at 2.38 eV. In both the cases, the intensity of this
peak (peak I) is quite small. In the standard parameter spectrum (\emph{cf.}
Fig.\ref{fig:acene6}a), the second peak corresponding to the to the
$1B_{3u}^{+}$ state is preceded by a weak shoulder (II$_{y}$) which
is due to a $y$-polarized photon, and has been identified as $2B_{2u}^{+}$
state.\vspace*{1cm}

In the screened parameter spectrum, this state is almost degenerate
with $1B_{3u}^{+}$ state and is part of the main feature II in the
spectrum. However, with both sets of parameters, the intensity is
mainly due to the $x$-polarized photon (see Tables \ref{acene6-mrsd4-tab}
and \ref{acene6-scr-mrsd4}). 

The remarkable feature of the dipole forbidden $1B_{3u}^{-}$ state
for hexacene is that its wave function mainly consists of double excitations,
with the single excitations making smaller contributions. This is
quite unlike the smaller acenes in which the $1B_{3u}^{-}$ state
consists mainly of singly excited configurations. Thus for this state,
contribution of correlation effects appears to increase with size
of the oligoacene. 

Angliker \emph{et al}.\cite{hexa-1} have studied the spectra of hexacene
experimentally as well as theoretically. They dissolved the hexacene
in the silicone oil and measured the $1B_{2u}^{+}$, $1B_{3u}^{-}$,
and $1B_{3u}^{+}$ excited states at 1.91 eV, 2.67 eV, and 3.94 eV
respectively. Theoretically they computed these states using P-P-P-SCF-SCI
method at 2.18 eV, 3.34 eV, and 4.05 eV respectively. Our correlated
calculations using standard parameters computed these states at 2.71
eV, 3.07 eV, and 4.61 eV respectively, while screened parameters found
these states to be at 2.38 eV, 2.77 eV, and 4.07 eV, respectively.
Thus, generally our standard parameter calculations overestimate the
experimental results of Angliker \emph{et al}.\cite{hexa-1} for $1B_{3u}^{-}$
state. However, with the screened parameters, our results are in good
agreement with their experimental results\cite{hexa-1} for the $1B_{3u}^{-}$
and $1B_{3u}^{+}$ excited states. Our screened parameter results
for the $1B_{2u}^{+}$ and the $1B_{3u}^{+}$ states also agree well
with their calculated results.\cite{hexa-1} Biermann \textit{et al.}\cite{biermann}
observed $1B_{3u}^{-}$ at 2.80 eV and $1B_{2u}^{+}$ at 1.90 eV.
Thus, in very good agreement with our screened parameter results for
the first DF $x$-polarized state, while slightly overestimated for
the first $y$-polarized state. Again, the trend is clear that the
results obtained using the screened parameters are in better agreement
with the experiments.\cite{hexa-1} 

Recently, Grimme and Parac\cite{grimme} used TDDFT and reported $1B_{3u}^{-}$
and $1B_{2u}^{+}$ states at 2.87 eV and 2.02 eV, which is in close
agreement with our screened parameter results. Heinze \emph{et al}.\cite{heinze113-hepta}
calculated the low-lying excited states of polyacenes using TDDFT
based coupled Kohn-Sham (CKS) methodology. They report $1B_{2u}^{+}$,
$1B_{3u}^{+}$, and $1B_{3u}^{-}$ states to be at 1.25 eV, 3.64 eV,
and 2.76 eV, respectively. Rubio and coworkers\cite{rubio} reported
these states to be at 1.50, 3.94, and 3.01 eV respectively. Except
for the $1B_{2u}^{+}$ state, our results of screened parameters calculations
are in very good agreement with their results. Additionally, Houk
\emph{et al.}\cite{houk-hexa} computed the HOMO to LUMO gap ($1B_{2u}^{+}$
state in our notations) at 1.54 eV, using the TDDFT approach. The
lower value of $1B_{2u}^{+}$ excitation energy from the TDDFT calculations
could be due to the usual problem of underestimating the band gaps,
associated with the DFT based approaches.

\subsubsection{Heptacene}

The final member of the oligoacene family which we have studied is
heptacene. It is a big molecule with seven diffused benzene rings,
and 30 $\pi$-electrons. It is an inherently unstable molecule,\cite{xu85-hepta}
whose synthesis has been under controversy.\cite{bendikov-hepta}
Initially, Clar\cite{clat-hepta} and Marschalk\cite{marschalk-hepta}
claimed the preparation of heptacene successfully. Later, Bailey and
Liaio reported the only other successful synthesis of heptacene.\cite{bailey-hepta}
Just after two years of the report of Bailey and Liaio, Clar retracted
his earlier claim,\cite{clat-hepta} and now it is commonly agreed
that heptacene cannot be prepared in the pure state.\cite{boggiano-hepta,notario102-hepta}
Therefore, no experimental data is available till date for the low-lying
excited states of heptacene. However, the low-lying excited states
have been studied theoretically by a few researchers using the TDDFT
method.\cite{houk-hexa,heinze113-hepta} 

\begin{figure}[H]
\begin{centering}
\includegraphics[width=10cm,keepaspectratio]{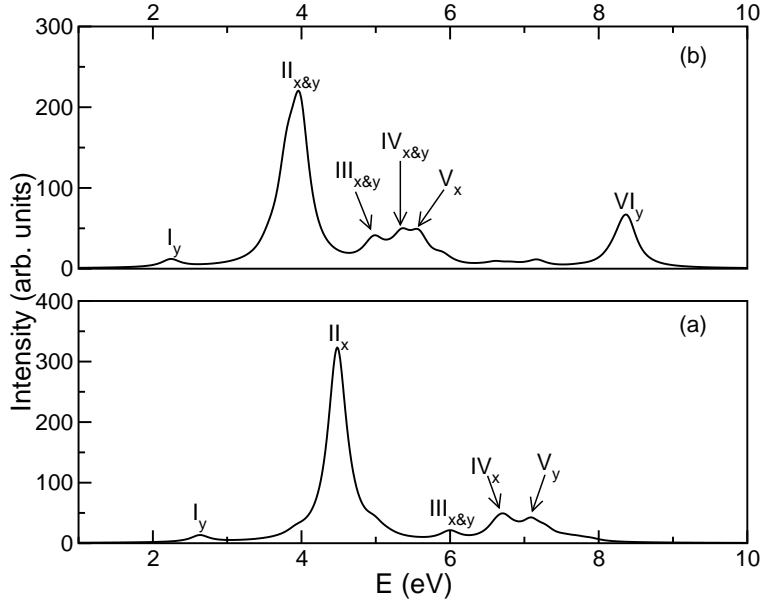}
\par\end{centering}

\caption{Linear optical absorption spectrum of heptacene computed with (a)
standard parameters and (b) screened parameters. MRSDCI method, coupled
with the P-P-P model Hamiltonian were used for the purpose.}

\label{fig:acene7} 
\end{figure}

We computed the linear absorption spectra of heptacene using standard
and screened parameters in the P-P-P model Hamiltonian, together with
the use of MRSDCI technique for the inclusion of the correlation effects.
The calculated spectra are presented in Fig.\ref{fig:acene7}, while
Tables \ref{acene7-mrsd4} and \ref{acene7-scr-mrsd4-tab} in the
appendix provide the detailed information regarding the excited state
properties. We note that, similar to the case of hexacene, the excited
state wave functions of the $1B_{3u}^{+}$ and $1B_{3u}^{-}$ states
are fundamentally different in that while the $1B_{3u}^{+}$ consists
mainly of one-particle excitations, while the $1B_{3u}^{-}$ state
is dominated by a double excitation, irrespective of the choice of
the Coulomb parameters. 

The first feature of Fig.\ref{fig:acene7}, corresponding to the $1B_{2u}^{+}$
state, was obtained at 2.63 eV from the standard parameter calculations,
and at 2.24 eV on using the screened parameters in the P-P-P Hamiltonian.
The dipole forbidden $1B_{3u}^{-}$ state was obtained at 2.73 eV
from the standard parameters, and at 2.35 eV from the screened parameters.
By both the parameters $1B_{3u}^{-}$ is at higher energy than $1B_{2u}^{+}$,
as is the case from tetracene onwards. Heinze \emph{et al}.\cite{heinze113-hepta}
also calculated the excitation energies for heptacene using the TDDFT
based CKS methodology.\cite{heinze113-hepta} They predicted $1B_{2u}^{+}$
state at 0.943 eV, $1B_{3u}^{-}$ state at 2.349 eV and $1B_{3u}^{+}$
state at 3.363 eV. Additionally, Houk \emph{et al}.\cite{houk-hexa}
used TDDFT method and computed the $1B_{2u}^{+}$ state (HOMO to LUMO
gap) at 1.24 eV. Our $1B_{2u}^{+}$ state is higher in comparison
to both the DFT based results. Lower values of energy for $1B_{2u}^{+}$
state obtained by DFT based methodology could be due to their well
known problem of underestimating the band gaps. The dipole forbidden
$1B_{3u}^{-}$ states computed by us, using screened parameters is
in perfect agreement with the results of Heinze \emph{et al}.,\cite{heinze113-hepta}
while the visible $1B_{3u}^{+}$ state with maximum intensity is obtained
at higher energies by both standard and screened parameters.

\section{Conclusions}

\label{sec:Conclusions}

In this paper we presented a large-scale correlated study of linear
optical absorption, and low-lying excited states, of oligoacenes ranging
from naphthalene to heptacene, using the P-P-P model Hamiltonian.
In order to investigate the influence of the Coulomb parameters on
the computed properties, these calculations were performed with two
sets of parameters in the P-P-P model: (i) standard Ohno parameters,\cite{ohno}
and (ii) a screened set of parameters.\cite{chandross} An extensive
comparison of our correlated results was made with: (i) the results
obtained using the single-particle Hückel model, (ii) the experimental
results, and (iii) the theoretical results of other authors. Next,
in a unified manner, we present the conclusions which can be drawn
from our calculations.

Qualitatively speaking, some aspects of the linear optical spectrum
computed using the Hückel model, were found to be very similar to
the one computed using the P-P-P model, for all the oligoacenes investigated.
For example, with both sets of calculations, the first peak of the
spectrum corresponded to the $1B_{2u}^{+}$ state, while the second,
and the most intense, peak of the spectrum corresponded to the $1B_{3u}^{+}$
state for all the acenes. Both sets of calculations also predicted
diminishing relative intensity of the $1B_{2u}^{+}$, with the increasing
conjugation length. Additionally, irrespective of the parameters used
in the P-P-P model, or the size of the acene involved, the many-particle
wave function of the $1B_{2u}^{+}$ state in all the cases consisted
mainly of the $H\rightarrow L$ excitation, in excellent agreement
with the Hückel model results. However, quantitatively speaking, for
all the oligoacenes, the optical gaps computed using the H\"uckel
model were much smaller than their P-P-P, and experimental, counterparts.

As far as the relative position of the dipole forbidden $1B_{3u}^{-}$
excited state is concerned, with both the sets of Coulomb parameters
the following trends were observed: (i) for oligoacenes up to anthracene,
$1B_{3u}^{-}$ occurs below the $1B_{2u}^{+}$ state, however, (ii)
from tetracene onwards the relative ordering of the two states gets
reversed, and $1B_{2u}^{+}$ state begins to occur below $1B_{3u}^{-}$.
This theoretical result was found to be in good agreement with most
of the experiments. Additionally, the influence of the electron correlation
effects on $1B_{3u}^{-}$ state appears to get stronger with the increasing
size of the acene involved. This manifests itself in form of the trend
that the contribution of the singly excited configurations to the
many-particle wave function of $1B_{3u}^{-}$ decreases with the increasing
conjugation length, and is eventually overshadowed by the doubly-excited
configurations for the larger acenes.

As mentioned earlier, for all the oligomers studied, irrespective
of the Coulomb parameters used in the P-P-P Hamiltonian, $1B_{3u}^{+}$
state was found to be the most intense peak of the spectrum, in excellent
agreement with all the available experiments. Additionally, unlike
$1B_{3u}^{-}$ state, for all the acenes, the many-particle wave function
of this most intense peak of the was dominated mainly by the singly-excited
configurations. Although, with the increasing conjugation length,
the relative contribution of the doubly-excited configurations to
the $1B_{3u}^{+}$ appears to increase. Therefore, it is conceivable
that for conjugation lengths longer than the ones considered here,
the contributions of the double excitations to $1B_{3u}^{+}$ could
overshadow those of the single excitations. Yet, for a given conjugation
length, the electron-correlation effects as judged from the relative
contribution of the double excitations, appear to be stronger in case
of $1B_{3u}^{-}$ state, as compared to the $1B_{3u}^{+}$ state.

Finally, we discuss the issue pertaining to the choice of Coulomb
parameters for the P-P-P model, when it comes to describing the linear
optical properties of oligoacenes. We notice that in an overall sense,
with the increasing conjugation length, the results obtained using
the screened parameters are in a better quantitative agreement with
the experiments as compared to those obtained using the standard parameters.
There were some acenes for which both the parameters gave reasonable
results, with the screened parameters energies slightly lower than
the experimental energies, and the standard parameters ones slightly
higher. Recalling that the screened parameters are generally used
to describe the solid-state or solvent effects (interchain screening)
, the better quantitative agreement obtained using the screened parameters
for the longer acenes, in our opinion, suggests the increasing importance
of the solid-state effects on longer acenes. 

In this paper we restricted ourselves to the low-lying excited states
of polyacenes which contribute to their linear optical properties.
However, it will also be interesting to explore the nature of their
two-photon states, which will contribute to the nonlinear, as well
as excited-state absorption, in these materials. Given the inherent
anisotropy (short-axis \textit{vs.} long-axis) of these materials,
several types of intermediate states will possibly govern their nonlinear
optical response. At present, studies along these directions are underway
in our group.

\begin{acknowledgments}
We thank Department of Science and Technology (DST), Government of
India, for providing financial support for this work under grant no.
SP/S2/M-10/2000. 
\end{acknowledgments}
\appendix

\section{Excited State properties}

\label{sec:appa}

Here we present the tables summarizing the results of our CI calculations
for various oligoacenes. The data presented in the tables includes
important configurations contributing to the many-body wave functions
of various excited states, their excitation energies, and transition
dipoles connecting them to the ground state. The results are presented
in separate subsections corresponding to each oligoacene, and include
calculations performed both with the standard parameters, and the
screened set of parameters in the P-P-P model.

\subsection{Naphthalene}

\begin{table}[H]

\caption{Excited states contributing to the linear absorption spectrum of
naphthalene computed using the FCI method coupled with the standard
parameters in the P-P-P model Hamiltonian. The table includes many
particle wave functions, excitation energies, and dipole matrix elements
of various states with respect to the ground state. DF corresponds
to dipole forbidden state. `$+c.c.$' indicates that the coefficient
of charge conjugate of a given configuration has the same sign, while
`$-c.c.$' implies that the two coefficients have opposite signs.}

\vspace{0.25cm}

\begin{centering}
\begin{tabular}{|c|c|c|c|c|}
\hline 
Peak&
State&
E (eV)&
Transition&
Wave Functions\tabularnewline
&
&
&
Dipole (\AA )&
\tabularnewline
\hline 
DF&
$1B_{3u}^{-}$&
3.61&
0.000&
$|H\rightarrow L+1\left\rangle \right.+c.c.(0.6235)$\tabularnewline
&
&
&
&
$|H\rightarrow L+1;H-3\rightarrow L\left\rangle \right.+c.c.(0.1345)$\tabularnewline
&
&
&
&
$|H-2\rightarrow L+3\left\rangle \right.+c.c.(0.1107)$\tabularnewline
\hline 
I&
$1B_{2u}^{+}$&
4.45&
0.551&
$|H\rightarrow L\left\rangle \right.(0.8773)$\tabularnewline
&
&
&
&
$|H-1\rightarrow L+1\left\rangle \right.(0.3521)$\tabularnewline
\hline 
II&
$1B_{3u}^{+}$&
5.99&
1.638&
$|H\rightarrow L+1\left\rangle \right.-c.c.(0.6524)$\tabularnewline
&
&
&
&
$|H\rightarrow L+1;H-3\rightarrow L\left\rangle \right.-c.c.(0.1015)$\tabularnewline
&
$2B_{2u}^{+}$&
6.10&
0.739&
$|H-1\rightarrow L+1\left\rangle \right.(0.8011)$\tabularnewline
&
&
&
&
$|H\rightarrow L\left\rangle \right.(0.2958)$\tabularnewline
&
&
&
&
$|H-2\rightarrow L+2\left\rangle \right.(0.2952)$\tabularnewline
\hline 
III&
$3B_{2u}^{+}$&
7.78&
1.028&
$|H-2\rightarrow L+2\left\rangle \right.(0.8155)$\tabularnewline
&
&
&
&
$|H-1\rightarrow L+1\left\rangle \right.(0.3002)$\tabularnewline
\hline 
IV&
$3B_{3u}^{+}$&
8.73&
0.281&
$|H-2\rightarrow L+3\left\rangle \right.-c.c.(0.4965)$\tabularnewline
&
&
&
&
$|H\rightarrow L+4\left\rangle \right.-c.c.(0.2280)$\tabularnewline
&
&
&
&
$|H\rightarrow L+1;H-1\rightarrow L+2\left\rangle \right.-c.c.(0.2260)$\tabularnewline
\hline
\end{tabular}\label{tab-acene2-fci} 
\par\end{centering}
\end{table}

\begin{table}[H]

\caption{Excited states contributing to the linear absorption spectrum of
naphthalene computed using the FCI method coupled with the screened
parameters in the P-P-P model Hamiltonian. The table includes many
particle wave functions, excitation energies, and dipole matrix elements
of various states with respect to the ground state. DF corresponds
to dipole forbidden state. `$+c.c.$' indicates that the coefficient
of charge conjugate of a given configuration has the same sign, while
`$-c.c.$' implies that the two coefficients have opposite signs.}

\vspace{0.25cm}

\begin{centering}
\begin{tabular}{|c|c|c|c|c|}
\hline 
Peak&
State&
E (eV)&
Transition&
Wave Functions\tabularnewline
&
&
&
Dipole (\AA )&
\tabularnewline
\hline 
DF&
$1B_{3u}^{-}$&
3.22&
0.000&
$|H\rightarrow L+1\left\rangle \right.-c.c.(0.5965)$\tabularnewline
&
&
&
&
$|H\rightarrow L+1;H-3\rightarrow L\left\rangle \right.+c.c.(0.1564)$\tabularnewline
\hline 
I&
$1B_{2u}^{+}$&
4.51&
0.719&
$\mid H\rightarrow L\left\rangle \right.(0.9145)$\tabularnewline
\hline 
II&
$1B_{3u}^{+}$&
5.30&
1.640&
$|H\rightarrow L+1\left\rangle \right.+c.c.(0.6466)$\tabularnewline
\hline 
III&
$2B_{2u}^{+}$&
6.12&
0.842&
$|H-1\rightarrow L+1\left\rangle \right.(0.9012)$\tabularnewline
&
&
&
&
$|H\rightarrow L+1;H-1\rightarrow L+3\left\rangle \right.-c.c.(0.1334)$\tabularnewline
\hline 
IV&
$3B_{2u}^{+}$&
7.52&
0.844&
$|H-2\rightarrow L+2\left\rangle \right.(0.8496)$\tabularnewline
\hline
\end{tabular}\label{acene2-scr-fci-tab} 
\par\end{centering}
\end{table}

\subsection{Anthracene}

\begin{table}[H]

\caption{Excited states contributing to the linear absorption spectrum of
anthracene computed using the FCI method coupled with the standard
parameters in the P-P-P model Hamiltonian. The table includes many
particle wave functions, excitation energies, and dipole matrix elements
of various states with respect to the ground state. DF corresponds
to dipole forbidden state. `$+c.c.$' indicates that the coefficient
of charge conjugate of a given configuration has the same sign, while
`$-c.c.$' implies that the two coefficients have opposite signs.}

\vspace{0.25cm}

\begin{centering}
\begin{tabular}{|c|c|c|c|c|}
\hline 
Peak&
State&
E (eV)&
Transition&
Wave Functions\tabularnewline
&
&
&
Dipole (\AA )&
\tabularnewline
\hline 
DF&
$1B_{3u}^{-}$&
3.25&
0.000&
$|H\rightarrow L+1\left\rangle \right.-c.c.(0.5963)$\tabularnewline
&
&
&
&
$|H\rightarrow L+1;H-3\rightarrow L\left\rangle \right.-c.c.(0.1329)$\tabularnewline
&
&
&
&
$|H-2\rightarrow L+3\left\rangle \right.+c.c.(0.1268)$\tabularnewline
\hline 
I&
$1B_{2u}^{+}$&
3.66&
0.728&
$|H\rightarrow L\left\rangle \right.(0.8894)$\tabularnewline
&
&
&
&
$|H-1\rightarrow L+1\left\rangle \right.(0.2097)$\tabularnewline
\hline 
II&
$1B_{3u}^{+}$&
5.34&
2.040&
$|H\rightarrow L+1\left\rangle \right.+c.c.(0.6296)$\tabularnewline
\hline 
III&
$4B_{2u}^{+}$&
6.94&
1.069&
$|H-2\rightarrow L+2\left\rangle \right.(0.6612)$\tabularnewline
&
&
&
&
$|H-1\rightarrow L+1\left\rangle \right.(0.4093)$\tabularnewline
\hline 
IV&
$3B_{3u}^{+}$&
7.70&
0.436&
$|H-2\rightarrow L+3\left\rangle \right.-c.c.(0.4831)$\tabularnewline
&
&
&
&
$|H\rightarrow L;H\rightarrow L+2\left\rangle \right.-c.c.(0.2403)$\tabularnewline
\hline 
V&
$7B_{2u}^{+}$&
8.30&
0.769&
$|H-4\rightarrow L+4\left\rangle \right.(0.6955)$\tabularnewline
&
&
&
&
$|H-1\rightarrow L+5\left\rangle \right.+c.c.(0.2120)$\tabularnewline
\hline
\end{tabular}\label{acene3-fci-tab}
\par\end{centering}
\end{table}

\begin{table}[H]

\caption{Excited states contributing to the linear absorption spectrum of
anthracene computed using the FCI method coupled with the screened
parameters in the P-P-P model Hamiltonian. The table includes many
particle wave functions, excitation energies, and dipole matrix elements
of various states with respect to the ground state. DF corresponds
to dipole forbidden state. `$+c.c.$' indicates that the coefficient
of charge conjugate of a given configuration has the same sign, while
`$-c.c.$' implies that the two coefficients have opposite signs.}

\vspace{0.25cm}

\begin{centering}
\begin{tabular}{|c|c|c|c|c|}
\hline 
Peak&
State&
E (eV)&
Transition&
Wave Functions\tabularnewline
&
&
&
Dipole (\AA )&
\tabularnewline
\hline 
DF&
$1B_{3u}^{-}$&
2.91&
0.000&
$\mid H\rightarrow L+1\left\rangle \right.-c.c.(0.5670)$\tabularnewline
&
&
&
&
$\mid H\rightarrow L+1;H-3\rightarrow L\left\rangle \right.+c.c.(0.1500)$\tabularnewline
\hline 
I&
$1B_{2u}^{+}$&
3.55&
0.747&
$\mid H\rightarrow L\left\rangle \right.(0.8890)$\tabularnewline
\hline 
II&
$1B_{3u}^{+}$&
4.64&
2.019&
$\mid H\rightarrow L+1\left\rangle \right.+c.c.(0.6215)$\tabularnewline
\hline 
III&
$2B_{2u}^{+}$&
5.93&
0.613&
$\mid H-1\rightarrow L+1\left\rangle \right.(0.7680)$\tabularnewline
&
&
&
&
$|H\rightarrow L+4\left\rangle \right.-c.c.(0.2283)$\tabularnewline
&
$3B_{2u}^{+}$&
6.01&
0.669&
$|H\rightarrow L+4\left\rangle \right.-c.c.(0.5024)$\tabularnewline
&
&
&
&
$|H-1\rightarrow L+1\left\rangle \right.(0.3616)$\tabularnewline
&
&
&
&
$|H-2\rightarrow L+2\left\rangle \right.(0.2987)$\tabularnewline
\hline 
IV&
$2B_{3u}^{+}$&
6.63&
0.519&
$\mid H\rightarrow L+5\left\rangle +c.c.\right.(0.3598)$\tabularnewline
&
&
&
&
$\mid H\rightarrow L;H\rightarrow L+2\left\rangle +c.c.\right.(0.3170)$\tabularnewline
&
&
&
&
$|H-2\rightarrow L+3\left\rangle -c.c.\right.(0.3161)$\tabularnewline
\hline 
V&
$7B_{2u}^{+}$&
7.82&
0.744&
$|H-4\rightarrow L+4\left\rangle \right.(0.5739)$\tabularnewline
&
&
&
&
$|H-3\rightarrow L+3\left\rangle \right.(0.3848)$\tabularnewline
&
&
&
&
$|H\rightarrow L;H-1\rightarrow L+2\left\rangle -c.c.\right.(0.2012)$\tabularnewline
\hline
\end{tabular}\label{acene3-scr-fci-tab}
\par\end{centering}
\end{table}

\subsection{Tetracene}

\begin{table}[H]

\caption{Excited states contributing to the linear absorption spectrum of
tetracene computed using the QCI method for $A_{g}$ and $B_{2u}$
states, and the MRSDCI method for $B_{3u}$ states, coupled with the
standard parameters in the P-P-P model Hamiltonian. The table includes
many particle wave functions, excitation energies, and dipole matrix
elements of various states with respect to the ground state. DF corresponds
to dipole forbidden state. `$+c.c.$' indicates that the coefficient
of charge conjugate of a given configuration has the same sign, while
`$-c.c.$' implies that the two coefficients have opposite signs. }

\vspace{0.25cm}

\begin{centering}
\begin{tabular}{|c|c|c|c|c|}
\hline 
Peak&
State&
E (eV)&
Transition&
Wave Functions\tabularnewline
&
&
&
Dipole (\AA )&
\tabularnewline
\hline 
DF&
$1B_{3u}^{-}$&
3.22&
0.000&
$|H\rightarrow L+2\left\rangle \right.-c.c.(0.5887)$\tabularnewline
&
&
&
&
$|H-1\rightarrow L+4\left\rangle \right.+c.c.(0.1395)$\tabularnewline
&
&
&
&
$|H\rightarrow L+2;H-4\rightarrow L\left\rangle \right.-c.c.(0.1289)$\tabularnewline
\hline 
I&
$1B_{2u}^{+}$&
3.16&
0.806&
$|H\rightarrow L\left\rangle \right.(0.8761)$\tabularnewline
\hline 
II&
$1B_{3u}^{+}$&
5.15&
2.392&
$|H\rightarrow L+2\left\rangle \right.+c.c.(0.6198)$\tabularnewline
\hline 
III&
$4B_{2u}^{+}$&
6.35&
0.799&
$|H-2\rightarrow L+2\left\rangle \right.(0.5623)$\tabularnewline
&
&
&
&
$|H-1\rightarrow L+1\left\rangle \right.(0.4056)$\tabularnewline
&
&
&
&
$|H-3\rightarrow L+3\left\rangle \right.(0.2211)$\tabularnewline
&
&
&
&
$|H-1\rightarrow L+5\left\rangle \right.+c.c.(0.2066)$\tabularnewline
\hline 
IV&
$9B_{2u}^{+}$&
7.90&
0.939&
$|H-3\rightarrow L+3\left\rangle \right.(0.4981)$\tabularnewline
&
&
&
&
$|H-2\rightarrow L+6\left\rangle \right.-c.c.(0.2665)$\tabularnewline
&
&
&
&
$|H-2\rightarrow L+2\left\rangle \right.(0.1925)$\tabularnewline
&
&
&
&
$|H-5\rightarrow L+5\left\rangle \right.(0.1921)$\tabularnewline
&
&
&
&
$|H-4\rightarrow L+4\left\rangle \right.(0.1879)$\tabularnewline
\hline
\end{tabular}\label{acene4-mrsd3-qci}
\par\end{centering}
\end{table}

\begin{table}[H]

\caption{Excited states contributing to the linear absorption spectrum of
tetracene computed using the QCI method for $A_{g}$ and $B_{2u}$
states, and the MRSDCI method for $B_{3u}$ states, coupled with the
screened parameters in the P-P-P model Hamiltonian. The table includes
many particle wave functions, excitation energies, and dipole matrix
elements of various states with respect to the ground state. DF corresponds
to dipole forbidden state. `$+c.c.$' indicates that the coefficient
of charge conjugate of a given configuration has the same sign, while
`$-c.c.$' implies that the two coefficients have opposite signs. }

\vspace{0.25cm}

\begin{centering}
\begin{tabular}{|c|c|c|c|c|}
\hline 
Peak&
State&
E (eV)&
Transition&
Wave Functions\tabularnewline
&
&
&
Dipole (\AA )&
\tabularnewline
\hline 
DF&
$1B_{3u}^{-}$&
3.02&
0.000&
$|H\rightarrow L+2\left\rangle \right.-c.c.(0.5750)$\tabularnewline
&
&
&
&
$|H-2\rightarrow L;H\rightarrow L+4\left\rangle \right.-c.c.(0.1552)$\tabularnewline
\hline 
I&
$1B_{2u}^{+}$&
2.97&
0.799&
$\mid H\rightarrow L\left\rangle \right.(0.8683)$\tabularnewline
\hline 
II&
$1B_{3u}^{+}$&
4.68&
2.356&
$\mid H\rightarrow L+2\left\rangle +c.c.(0.6269)\right.$\tabularnewline
\hline 
III&
$3B_{2u}^{+}$&
5.41&
0.671&
$\mid H-1\rightarrow L+1\left\rangle \right.(0.7657)$\tabularnewline
&
&
&
&
$\mid H\rightarrow L+3\left\rangle -c.c.(0.2076)\right.$\tabularnewline
\hline 
IV&
$2B_{3u}^{+}$&
6.03&
0.573&
$\mid H\rightarrow L;H\rightarrow L+1\left\rangle +c.c.\right.(0.4998)$\tabularnewline
&
&
&
&
$\mid H-4\rightarrow L+1\left\rangle \right.+c.c.(0.2119)$\tabularnewline
&
&
&
&
$\mid H-1\rightarrow L+1;H\rightarrow L+1\left\rangle +c.c.\right.(0.2004)$\tabularnewline
&
$4B_{2u}^{+}$&
6.10&
0.841&
$\mid H-2\rightarrow L+2\left\rangle \right.(0.7957)$\tabularnewline
\hline 
V&
$5B_{3u}^{+}$&
7.22&
0.413&
$\mid H-3\rightarrow L+2\left\rangle -c.c.\right.(0.3768)$\tabularnewline
&
&
&
&
$\mid H\rightarrow L;H\rightarrow L+1\left\rangle -c.c.\right.(0.2426)$\tabularnewline
&
&
&
&
$\mid H\rightarrow L;H\rightarrow L+5\left\rangle +c.c.\right.(0.2345)$\tabularnewline
&
&
&
&
$\mid H-1\rightarrow L+4\left\rangle +c.c.\right.(0.2177)$\tabularnewline
&
$9B_{2u}^{+}$&
7.37&
0.901&
$\mid H-3\rightarrow L+3\left\rangle \right.(0.7722)$\tabularnewline
\hline 
VI&
$16B_{2u}^{+}$&
8.44&
0.650&
$\mid H-5\rightarrow L+5\left\rangle \right.(0.6119)$\tabularnewline
&
&
&
&
$\mid H\rightarrow L+1;H-4\rightarrow L+1\left\rangle +c.c.\right.(0.1790)$\tabularnewline
\hline
\end{tabular}\label{acene4-scr-mrsd3-qci}
\par\end{centering}
\end{table}

\subsection{Pentacene}

\begin{table}[H]

\caption{Excited states contributing to the linear absorption spectrum of
pentacene computed using the QCI method for $A_{g}$ and $B_{2u}$
states, and the MRSDCI method for $B_{3u}$ states, coupled with the
standard parameters in the P-P-P model Hamiltonian. The table includes
many particle wave functions, excitation energies, and dipole matrix
elements of various states with respect to the ground state. DF corresponds
to dipole forbidden state. `$+c.c.$' indicates that the coefficient
of charge conjugate of a given configuration has the same sign, while
`$-c.c.$' implies that the two coefficients have opposite signs. }

\vspace{0.25cm}

\begin{centering}
\begin{tabular}{|c|c|c|c|c|}
\hline 
Peak&
State&
E (eV)&
Transition&
Wave Functions\tabularnewline
&
&
&
Dipole (\AA )&
\tabularnewline
\hline 
DF &
$1B_{3u}^{-}$&
3.17&
0.000&
$|H\rightarrow L+2\left\rangle +c.c.\right.(0.5519)$\tabularnewline
&
&
&
&
$|H\rightarrow L;H\rightarrow L+1\left\rangle +c.c.\right.(0.1439)$\tabularnewline
&
&
&
&
$|H-1\rightarrow L+4\left\rangle +c.c.\right.(0.1341)$\tabularnewline
\hline 
I&
$1B_{2u}^{+}$&
2.86&
0.838&
$|H\rightarrow L\left\rangle \right.(0.8568)$\tabularnewline
\hline 
II&
$1B_{3u}^{+}$&
5.01&
2.721&
$|H\rightarrow L+2\left\rangle -c.c.\right.(0.5950)$\tabularnewline
&
&
&
&
$|H-1\rightarrow L+4\left\rangle -c.c.\right.(0.1467)$\tabularnewline
&
$3B_{2u}^{+}$&
5.16&
0.432&
$|H-1\rightarrow L+1\left\rangle \right.(0.5108)$\tabularnewline
&
&
&
&
$|H\rightarrow L+3\left\rangle +c.c.\right.(0.3155)$\tabularnewline
&
&
&
&
$|H-2\rightarrow L+2\left\rangle \right.(0.2668)$\tabularnewline
\hline 
III&
$4B_{3u}^{+}$&
7.09&
0.462&
$|H-1\rightarrow L+4\left\rangle -c.c.\right.(0.3674)$\tabularnewline
&
&
&
&
$|H\rightarrow L;H\rightarrow L+5\left\rangle +c.c.\right.(0.2329)$\tabularnewline
&
&
&
&
$|H\rightarrow L;H\rightarrow L+1\left\rangle -c.c.\right.(0.1848)$\tabularnewline
&
$5B_{3u}^{+}$&
7.23&
0.413&
$|H-1\rightarrow L+1;H\rightarrow L+1\left\rangle -c.c.\right.(0.2762)$\tabularnewline
&
&
&
&
$|H\rightarrow L;H-1\rightarrow L+3\left\rangle -c.c.\right.(0.2266)$\tabularnewline
&
&
&
&
$|H\rightarrow L+7\left\rangle +c.c.\right.(0.2263)$\tabularnewline
&
$11B_{2u}^{+}$&
7.27&
0.668&
$|H-3\rightarrow L+3\left\rangle \right.(0.4121)$\tabularnewline
&
&
&
&
$|H-6\rightarrow L+3\left\rangle +c.c.\right.(0.2475)$\tabularnewline
&
&
&
&
$|H-4\rightarrow L+4\left\rangle \right.(0.2374)$\tabularnewline
\hline 
IV&
$12B_{2u}^{+}$&
7.58&
0.672&
$|H-3\rightarrow L+3\left\rangle \right.(0.3294)$\tabularnewline
&
&
&
&
$|H-2\rightarrow L+2\left\rangle \right.(0.2682)$\tabularnewline
&
&
&
&
$|H-5\rightarrow L+5\left\rangle \right.(0.2125)$\tabularnewline
&
&
&
&
$|H-1\rightarrow L+5\left\rangle -c.c.\right.(0.2060)$\tabularnewline
&
$6B_{3u}^{+}$&
7.60&
0.495&
$|H-3\rightarrow L+2\left\rangle -c.c.\right.(0.3907)$\tabularnewline
&
&
&
&
$|H-1\rightarrow L+4\left\rangle -c.c.\right.(0.2922)$\tabularnewline
&
&
&
&
$|H\rightarrow L;H\rightarrow L+1\left\rangle +-c.c.\right.(0.2172)$\tabularnewline
\hline
\end{tabular}\label{acene5-mrsd3-qci-tab}
\par\end{centering}
\end{table}

\begin{table}[H]

\caption{Excited states contributing to the linear absorption spectrum of
pentacene computed using the QCI method for $A_{g}$ and $B_{2u}$
states, and the MRSDCI method for $B_{3u}$ states, coupled with the
screened parameters in the P-P-P model Hamiltonian. The table includes
many particle wave functions, excitation energies, and dipole matrix
elements of various states with respect to the ground state. DF corresponds
to dipole forbidden state. `$+c.c.$' indicates that the coefficient
of charge conjugate of a given configuration has the same sign, while
`$-c.c.$' implies that the two coefficients have opposite signs. }

\vspace{0.25cm}

\begin{centering}
\begin{tabular}{|c|c|c|c|c|}
\hline 
Peak&
State&
E (eV)&
Transition&
Wave Functions\tabularnewline
&
&
&
Dipole (\AA )&
\tabularnewline
\hline 
DF&
$1B_{3u}^{-}$&
2.99&
0.000&
$\mid H\rightarrow L+2\left\rangle \right.+c.c.(0.5117)$\tabularnewline
&
&
&
&
$\mid H\rightarrow L;H\rightarrow L+1\left\rangle \right.-c.c.(0.2348)$\tabularnewline
&
&
&
&
$\mid H\rightarrow L+2;H-4\rightarrow L\left\rangle \right.-c.c.(0.1394)$\tabularnewline
\hline 
I&
$1B_{2u}^{+}$&
2.65&
0.824&
$\mid H\rightarrow L\left\rangle \right.(0.8464)$\tabularnewline
\hline 
II&
$1B_{3u}^{+}$&
4.65&
2.642&
$\mid H\rightarrow L+2\left\rangle \right.-c.c.(0.6146)$\tabularnewline
&
$2B_{2u}^{+}$&
4.63&
0.317&
$\mid H\rightarrow L+3\left\rangle \right.+c.c.(0.5123)$\tabularnewline
&
&
&
&
$\mid H\rightarrow L\left\rangle \right.(0.3994)$\tabularnewline
\hline 
III&
$4B_{3u}^{+}$&
6.47&
0.593&
$\mid H-4\rightarrow L+1\left\rangle -c.c.\right.(0.4034)$\tabularnewline
&
&
&
&
$|H-1\rightarrow L+1;H\rightarrow L+1\left\rangle +c.c.\right.(0.2226)$\tabularnewline
&
&
&
&
$|H\rightarrow L;H-1\rightarrow L+3\left\rangle -c.c.\right.(0.2216)$\tabularnewline
\hline 
IV&
$11B_{2u}^{+}$&
6.77&
0.782&
$\mid H-3\rightarrow L+3\left\rangle \right.(0.7460)$\tabularnewline
&
&
&
&
$\mid H\rightarrow L;H\rightarrow L;H-3\rightarrow L+3\left\rangle \right.(0.2030)$\tabularnewline
&
&
&
&
$\mid H\rightarrow L+6\left\rangle \right.-c.c.(0.1413)$\tabularnewline
\hline
\end{tabular}\label{acene5-scr-mrsd3-qci}
\par\end{centering}
\end{table}

\subsection{Hexacene}

\begin{table}[H]

\caption{Excited states contributing to the linear absorption spectrum of
hexacene computed using the MRSDCI method, coupled with the standard
parameters in the P-P-P model Hamiltonian. The table includes many-particle
wave functions, excitation energies, and dipole matrix elements of
various states with respect to the ground state. DF corresponds to
dipole forbidden state. `$+c.c.$' indicates that the coefficient
of charge conjugate of a given configuration has the same sign, while
`$-c.c.$' implies that the two coefficients have opposite signs. }

\vspace{0.25cm}

\begin{centering}
\begin{tabular}{|c|c|c|c|c|}
\hline 
Peak&
State&
E (eV)&
Transition&
Wave Functions\tabularnewline
&
&
&
Dipole (\AA )&
\tabularnewline
\hline 
DF&
$1B_{3u}^{-}$&
3.07&
0.000&
$\mid H\rightarrow L;H\rightarrow L+1\left\rangle \right.+c.c.(0.4407)$\tabularnewline
&
&
&
&
$\mid H\rightarrow L+3\left\rangle \right.+c.c.(0.2729)$\tabularnewline
&
&
&
&
$\mid H\rightarrow L+7\left\rangle \right.+c.c.(0.1758)$\tabularnewline
\hline 
I&
$1B_{2u}^{+}$&
2.71&
0.812&
$\mid H\rightarrow L\left\rangle \right.(0.8584)$\tabularnewline
\hline 
II&
$2B_{2u}^{+}$&
4.19&
0.733&
$\mid H\rightarrow L+2\left\rangle +c.c.\right.(0.5045)$\tabularnewline
&
&
&
&
$\mid H\rightarrow L;H-1\rightarrow L+3\left\rangle \right.+c.c.(0.1810)$\tabularnewline
&
&
&
&
$\mid H\rightarrow L;H\rightarrow L+4\left\rangle \right.-c.c.(0.1720)$\tabularnewline
\hline 
III&
$1B_{3u}^{+}$&
4.61&
3.049&
$\mid H\rightarrow L+3\left\rangle \right.-c.c.(0.5550)$\tabularnewline
&
&
&
&
$\mid H-4\rightarrow L+1\left\rangle \right.+c.c.(0.1653)$\tabularnewline
\hline 
IV&
$5B_{2u}^{+}$&
5.87&
0.764&
$\mid H\rightarrow L+6\left\rangle \right.+c.c.(0.4457)$\tabularnewline
&
&
&
&
$\mid H-1\rightarrow L+5\left\rangle \right.+c.c.(0.1821)$\tabularnewline
\hline 
V&
$4B_{3u}^{+}$&
6.42&
0.549&
$\mid H\rightarrow L;H\rightarrow L+1\left\rangle \right.+c.c.(0.3513)$\tabularnewline
&
&
&
&
$\mid H-1\rightarrow L+4\left\rangle \right.+c.c.(0.2390)$\tabularnewline
&
&
&
&
$\mid H-1\rightarrow L+1;H\rightarrow L+1\left\rangle \right.-c.c.(0.2092)$\tabularnewline
&
$9B_{2u}^{+}$&
6.51&
0.619&
$\mid H-1\rightarrow L+5\left\rangle \right.+c.c.(0.2787)$\tabularnewline
&
&
&
&
$\mid H\rightarrow L+1;H\rightarrow L+3\left\rangle \right.+c.c.(0.2784)$\tabularnewline
\hline 
VI&
$11B_{2u}^{+}$&
6.87&
0.256&
$\mid H-3\rightarrow L+3\left\rangle \right.(0.4731)$\tabularnewline
&
&
&
&
$\mid H-1\rightarrow L+5\left\rangle \right.+c.c.(0.2414)$\tabularnewline
&
&
&
&
$\mid H-5\rightarrow L+5\left\rangle \right.(0.2177)$\tabularnewline
&
$6B_{3u}^{+}$&
6.90&
0.478&
$\mid H\rightarrow L;H\rightarrow L+8\left\rangle \right.+c.c.(0.4054)$\tabularnewline
&
&
&
&
$\mid H\rightarrow L;H-1\rightarrow L+6\left\rangle \right.+c.c.(0.2608)$\tabularnewline
&
&
&
&
$\mid H\rightarrow L;H-2\rightarrow L+5\left\rangle \right.+c.c.(0.2090)$\tabularnewline
VI&
$7B_{3u}^{+}$&
7.01&
0.585&
$\mid H-1\rightarrow L+4\left\rangle \right.+c.c.(0.3445)$\tabularnewline
&
&
&
&
$\mid H\rightarrow L;H-1\rightarrow L+2\left\rangle \right.+c.c.(0.2463)$\tabularnewline
&
&
&
&
$\mid H\rightarrow L;H\rightarrow L+8\left\rangle \right.+c.c.(0.2163)$\tabularnewline
&
&
&
&
$\mid H-2\rightarrow L+3\left\rangle \right.+c.c.(0.2019)$\tabularnewline
\hline
\end{tabular}\label{acene6-mrsd4-tab}
\par\end{centering}
\end{table}

\begin{table}[H]

\caption{Excited states contributing to the linear absorption spectrum of
hexacene computed using the MRSDCI method, coupled with the screened
parameters in the P-P-P model Hamiltonian. The table includes many
particle wave functions, excitation energies, and dipole matrix elements
of various states with respect to the ground state. DF corresponds
to dipole forbidden state. `$+c.c.$' indicates that the coefficient
of charge conjugate of a given configuration has the same sign, while
`$-c.c.$' implies that the two coefficients have opposite signs. }

\vspace{0.25cm}

\begin{centering}
\begin{tabular}{|c|c|c|c|c|}
\hline 
Peak&
State&
E (eV)&
Transition&
Wave Functions\tabularnewline
&
&
&
Dipole (\AA )&
\tabularnewline
\hline 
DF&
$1B_{3u}^{-}$&
2.77&
0.000&
$\mid H\rightarrow L;H\rightarrow L+1\left\rangle \right.+c.c.(0.4943)$\tabularnewline
&
&
&
&
$\mid H\rightarrow L+3\left\rangle \right.+c.c.(0.1830)$\tabularnewline
&
&
&
&
$\mid H-1\rightarrow L;H\rightarrow L+2\left\rangle \right.-c.c.(0.1819)$\tabularnewline
\hline 
I&
$1B_{2u}^{+}$&
2.38&
0.787&
$\mid H\rightarrow L\left\rangle \right.(0.8683)$\tabularnewline
\hline 
II&
$2B_{2u}^{+}$&
3.94&
0.716&
$\mid H-1\rightarrow L+1\left\rangle \right.(0.7077)$\tabularnewline
&
&
&
&
$\mid H\rightarrow L+2\left\rangle \right.-c.c.(0.3154)$\tabularnewline
&
$1B_{3u}^{+}$&
4.07&
2.948&
$\mid H\rightarrow L+3\left\rangle -c.c.(0.5956)\right.$\tabularnewline
\hline 
III&
$6B_{2u}^{+}$&
5.58&
0.414&
$\mid H\rightarrow L+6\left\rangle +c.c.(0.3583)\right.$\tabularnewline
&
&
&
&
$\mid H-1\rightarrow L+5\left\rangle \right.-c.c.(0.3188)$\tabularnewline
&
&
&
&
$\mid H-2\rightarrow L+2\left\rangle \right.(0.2851)$\tabularnewline
&
$4B_{3u}^{+}$&
5.61&
0.971&
$\mid H-1\rightarrow L+4\left\rangle \right.+c.c.(0.4188)$\tabularnewline
&
&
&
&
$\mid H\rightarrow L+7\left\rangle -c.c.(0.2382)\right.$\tabularnewline
&
&
&
&
$\mid H\rightarrow L;H-1\rightarrow L+2\left\rangle +c.c.(0.2368)\right.$\tabularnewline
&
&
&
&
$\mid H-1\rightarrow L+1;H\rightarrow L+1\left\rangle -c.c.(0.2054)\right.$\tabularnewline
\hline 
IV&
$9B_{2u}^{+}$&
5.93&
0.672&
$\mid H-2\rightarrow L+2\left\rangle \right.(0.6354)$\tabularnewline
&
&
&
&
$\mid H-1\rightarrow L+5\left\rangle -c.c.(0.2745)\right.$\tabularnewline
\hline 
V&
$8B_{3u}^{+}$&
6.23&
0.555&
$\mid H-2\rightarrow L+3\left\rangle \right.+c.c.(0.5539)$\tabularnewline
&
&
&
&
$\mid H\rightarrow L+7\left\rangle -c.c.(0.1646)\right.$\tabularnewline
&
&
&
&
$\mid H\rightarrow L;H-1\rightarrow L+2\left\rangle +c.c.(0.1097)\right.$\tabularnewline
\hline 
VI&
$18B_{2u}^{+}$&
7.71&
0.714&
$\mid H-3\rightarrow L+3\left\rangle \right.(0.6973)$\tabularnewline
\hline
\end{tabular}\label{acene6-scr-mrsd4}
\par\end{centering}
\end{table}

\subsection{Heptacene}

\begin{table}[H]

\caption{Excited states contributing to the linear absorption spectrum of
heptacene computed using the MRSDCI method, coupled with the standard
parameters in the P-P-P model Hamiltonian. The table includes many
particle wave functions, excitation energies, and dipole matrix elements
of various states with respect to the ground state. DF corresponds
to dipole forbidden state. `$+c.c.$' indicates that the coefficient
of charge conjugate of a given configuration has the same sign, while
`$-c.c.$' implies that the two coefficients have opposite signs. }

\vspace{0.25cm}

\begin{centering}
\begin{tabular}{|c|c|c|c|c|}
\hline 
Peak&
State&
E (eV)&
Transition&
Wave Functions\tabularnewline
&
&
&
Dipole (\AA )&
\tabularnewline
\hline 
DF&
$1B_{3u}^{-}$&
2.73&
0.000&
$\mid H\rightarrow L;H\rightarrow L+1\left\rangle \right.+c.c.(0.4896)$\tabularnewline
&
&
&
&
$\mid H\rightarrow L+1;H-2\rightarrow L\left\rangle \right.+c.c.(0.1904)$\tabularnewline
&
&
&
&
$\mid H\rightarrow L+1;H\rightarrow L+2\left\rangle \right.+c.c.(0.1712)$\tabularnewline
\hline 
I&
$1B_{2u}^{+}$&
2.63&
0.793&
$\mid H\rightarrow L\left\rangle \right.(0.8563)$\tabularnewline
\hline 
II&
$1B_{3u}^{+}$&
4.48&
3.251&
$\mid H\rightarrow L+3\left\rangle \right.-c.c.(0.5033)$\tabularnewline
&
&
&
&
$\mid H\rightarrow L;H\rightarrow L+1\left\rangle -c.c.(0.2170)\right.$\tabularnewline
&
&
&
&
$\mid H-1\rightarrow L+5\left\rangle +c.c.(0.1789)\right.$\tabularnewline
\hline 
III&
$6B_{2u}^{+}$&
5.96&
0.326&
$\mid H\rightarrow L;H\rightarrow L+5\left\rangle -c.c.(0.3368)\right.$\tabularnewline
&
&
&
&
$\mid H\rightarrow L+1;H\rightarrow L+3\left\rangle +c.c.(0.3312)\right.$\tabularnewline
&
&
&
&
$\mid H-1\rightarrow L+1\left\rangle \right.(0.1997)$\tabularnewline
&
$4B_{3u}^{+}$&
6.01&
0.490&
$\mid H\rightarrow L;H-1\rightarrow L+2\left\rangle -c.c.(0.4204)\right.$\tabularnewline
&
&
&
&
$\mid H-2\rightarrow L+2;H\rightarrow L+1\left\rangle -c.c.(0.1781)\right.$\tabularnewline
\hline 
IV&
$7B_{3u}^{+}$&
6.70&
0.624&
$\mid H-1\rightarrow L+5\left\rangle +c.c.(0.3258)\right.$\tabularnewline
&
&
&
&
$\mid H\rightarrow L;H-1\rightarrow L+2\left\rangle \right.-c.c.(0.2972)$\tabularnewline
&
&
&
&
$\mid H\rightarrow L;H\rightarrow L+4\left\rangle \right.+c.c.(0.2444)$\tabularnewline
\hline 
V&
$12B_{2u}^{+}$&
7.09&
0.736&
$\mid H-2\rightarrow L+2\left\rangle \right.(0.3457)$\tabularnewline
&
&
&
&
$\mid H-3\rightarrow L+3\left\rangle \right.(0.3312)$\tabularnewline
\hline
\end{tabular}\label{acene7-mrsd4}
\par\end{centering}
\end{table}

\begin{table}[H]

\caption{Excited states contributing to the linear absorption spectrum of
heptacene computed using the MRSDCI method, coupled with the screened
parameters in the P-P-P model Hamiltonian. The table includes many
particle wave functions, excitation energies, and dipole matrix elements
of various states with respect to the ground state. DF corresponds
to dipole forbidden state. `$+c.c.$' indicates that the coefficient
of charge conjugate of a given configuration has the same sign, while
`$-c.c.$' implies that the two coefficients have opposite signs. }

\vspace{0.25cm}

\begin{centering}
\begin{tabular}{|c|c|c|c|c|}
\hline 
Peak&
State&
E (eV)&
Transition&
Wave Functions\tabularnewline
&
&
&
Dipole (\AA )&
\tabularnewline
\hline 
DF&
$1B_{3u}^{-}$&
2.35&
0.000&
$\mid H\rightarrow L;H\rightarrow L+1\left\rangle \right.+c.c.(0.5062)$\tabularnewline
&
&
&
&
$\mid H-1\rightarrow L;H\rightarrow L+2\left\rangle \right.+c.c.(0.1942)$\tabularnewline
&
&
&
&
$\mid H\rightarrow L+1;H\rightarrow L+2\left\rangle \right.+c.c.(0.1648)$\tabularnewline
\hline 
I&
$1B_{2u}^{+}$&
2.24&
0.795&
$\mid H\rightarrow L\left\rangle \right.(0.8675)$\tabularnewline
\hline 
II&
$1B_{3u}^{+}$&
3.80&
1.900&
$\mid H\rightarrow L;H\rightarrow L+1\left\rangle \right.-c.c.(0.4795)$\tabularnewline
&
&
&
&
$\mid H-1\rightarrow L+1;H\rightarrow L+1\left\rangle \right.-c.c.(0.2362)$\tabularnewline
&
&
&
&
$\mid H\rightarrow L+3\left\rangle \right.-c.c.(0.1781)$\tabularnewline
&
$3B_{2u}^{+}$&
3.87&
0.459&
$\mid H\rightarrow L+2\left\rangle \right.+c.c.(0.5254)$\tabularnewline
&
&
&
&
$\mid H-1\rightarrow L+1\left\rangle \right.(0.4140)$\tabularnewline
&
$2B_{3u}^{+}$&
3.98&
2.553&
$\mid H\rightarrow L+3\left\rangle \right.-c.c.(0.5710)$\tabularnewline
\hline 
III&
$4B_{2u}^{+}$&
4.92&
0.459&
$\mid H-2\rightarrow L+2\left\rangle \right.+(0.4342)$\tabularnewline
&
&
&
&
$\mid H-1\rightarrow L+4\left\rangle \right.+c.c.(0.4189)$\tabularnewline
&
&
&
&
$\mid H\rightarrow L+6\left\rangle \right.+c.c.(0.2618)$\tabularnewline
&
$4B_{3u}^{+}$&
4.98&
0.799&
$\mid H\rightarrow L;H-1\rightarrow L+2\left\rangle \right.-c.c.(0.4219)$\tabularnewline
\hline 
IV&
$5B_{3u}^{+}$&
5.34&
0.668&
$\mid H\rightarrow L+7\left\rangle \right.-c.c.(0.4384)$\tabularnewline
&
&
&
&
$\mid H\rightarrow L;H-1\rightarrow L+2\left\rangle \right.-c.c.(0.2542)$\tabularnewline
&
$7B_{2u}^{+}$&
5.36&
0.609&
$\mid H-2\rightarrow L+2\left\rangle \right.(0.5520)$\tabularnewline
&
&
&
&
$\mid H-1\rightarrow L+4\left\rangle +c.c.\right.(0.3475)$\tabularnewline
\hline 
V&
$7B_{3u}^{+}$&
5.57&
0.867&
$\mid H-1\rightarrow L+5\left\rangle \right.+c.c.(0.4741)$\tabularnewline
&
&
&
&
$\mid H\rightarrow L+7\left\rangle \right.-c.c.(0.2927)$\tabularnewline
&
&
&
&
$\mid H\rightarrow L;H-1\rightarrow L+2\left\rangle \right.-c.c.(0.1582)$\tabularnewline
\hline 
VI&
$20B_{2u}^{+}$&
8.39&
0.871&
$\mid H-4\rightarrow L+4\left\rangle \right.(0.7058)$\tabularnewline
\hline
\end{tabular}\label{acene7-scr-mrsd4-tab}
\par\end{centering}
\end{table}

\end{document}